\documentclass[12pt]{article}

\usepackage{graphicx,epsfig,rotate,amsmath}    
\usepackage{latexsym,color,amssymb,citesort}
\newcommand{\be}{\begin{equation}}
\newcommand{\ee}{\end{equation}}
\newcommand{\eq}{\begin{eqnarray}}
\newcommand{\en}{\end{eqnarray}}
\newcommand{\bea}{\begin{eqnarray}}
\newcommand{\eea}{\end{eqnarray}}

\newcommand{\ed}{\end{document}}
\newcommand{\bc}{\begin{center}}
\newcommand{\ec}{\end{center}}
\newcommand{\btau}{\mbox{\boldmath $\mathbf\tau$}}
\newcommand{\btheta}{\mbox{\boldmath $\mathbf\theta$}}

\begin{document}

\thispagestyle{empty}

\begin{center}

\vspace{1.75cm}
{\Large{\bf Three particles in a finite volume}}

\vspace{0.5cm}

\today

\vspace{0.5cm}

K.~Polejaeva and A.~Rusetsky

\vspace{2em}

\begin{tabular}{c}
{\it Helmholtz--Institut f\"ur Strahlen-- und Kernphysik}\\
{\it and Bethe Center for Theoretical Physics, Universit\"at Bonn}\\
{\it  D--53115 Bonn, Germany}
\end{tabular}

\end{center}

\vspace{1cm}

{\abstract
{Within the non-relativistic
 potential scattering theory, we derive a generalized
version of the L\"uscher formula, which includes three-particle
inelastic channels. Faddeev equations in a finite volume are
discussed in detail. It is proved that, even in the presence of
the three-particle intermediate states, the discrete spectrum
in a finite box  is determined by the infinite-volume elements
of the scattering $S$-matrix up to 
corrections, exponentially suppressed at large volumes.
}
}

\vskip1cm

{\footnotesize{\begin{tabular}{ll}
{\bf{Pacs:}}$\!\!\!\!$& 12.38.Gc, 11.10.St, 11.80.Jy\\
{\bf{Keywords:}}$\!\!\!\!$& Resonances in lattice QCD,
field theory in a finite volume, Faddeev equations 
\end{tabular}}
}
\clearpage


\section{Introduction}

The nature of the Roper resonance $N(1440)$ has been an open question for decades.
It is quite unnatural -- at least, from the point of view of the quark models~\cite{Isgur1,Isgur2} -- 
that its mass turns out to be lower than that of the negative-parity ground state
$N(1535)$. Different phenomenological approaches have been employed so far to
explain such a level ordering. For example, $N(1440)$ was assumed to be a hybrid
state with excited gluon field configurations~\cite{hybrid1,hybrid2}, a breathing mode of the
ground state~\cite{breather}, and a five-quark (meson-baryon) state~\cite{fivequark}. The properties of the Roper resonance have been studied
within the Skyrme model~\cite{ulf1} and the bag model~\cite{ulf2}.

Up to now, numerous calculations of the excited spectrum of the 
nucleon~\cite{Gockeler:2001db,Melnitchouk:2002eg,Lee:2002gn1,Lee:2002gn2,Edwards:2003cd,F.Lee,Sasaki:2003xc1,Sasaki:2003xc2,Sasaki:2003xc3,Basak:2007kj,Cohen:2009zk,Bulava:2010yg1,Bulava:2010yg2,Engel:2010my,Mahbub:2010rm,Lin:2011da} 
(see \textit{e.g.}, ref.~\cite{Lin:2011ti} for a recent state-of-art review)
have not resolved the issue. In particular, a reverse level ordering (the same as in the
constituent quark model) between 
$N(1440)$ and $N(1535)$ has been reported in some lattice calculations~\cite{F.Lee}, albeit
the situation there is far from being clear. It should be also pointed out that
a large chiral curvature for the Roper resonance mass is not compatible with the findings
of ref.~\cite{Borasoy:2006fk}, where the dependence of the Roper mass on the pion mass
was investigated within the framework of Chiral Perturbation Theory. 
According to these findings, the level crossing between $N(1440)$ and 
$N(1535)$ states, emerging as a result of chiral extrapolation to the small 
quark masses, does not constitute to a plausible scenario.  

There is, however, one issue that has not been addressed in all above investigations
so far. The $N(1440)$ is a resonance and not a stable state which would correspond
to an isolated energy level in lattice simulations. This, in particular, means that
the energy levels measured on the lattice will be volume-dependent and the true resonance
pole position should be extracted from the volume-dependent spectrum.

The case of the elastic low-lying resonances on the lattice has been investigated in detail.
Namely, the
L\"uscher formula~\cite{luescher-torus} enables one to uniquely relate the discrete energy
levels in a finite box to the elastic scattering phase shift in the infinite
volume, measured at the same energy. This eventually opens the way for
 the extraction of the parameters
of the elastic resonances -- their masses and widths -- in the lattice QCD 
(for illustration see, \textit{e.g.}, refs.~\cite{rho1,rho2}).

The case of the inelastic resonances, however, is more complicated. Albeit
the L\"uscher approach can be straightforwardly generalized to the case of the
coupled two-particle channels~\cite{lage-KN,He,scalar}, the disentanglement of physical
observables becomes a more delicate affair, since there is more than one observable
at a single energy. In order to circumvent this problem, in refs.~\cite{scalar,oset} the use
of the twisted boundary conditions or asymmetric boxes
 was advocated for the $\pi\pi-K\bar K$ coupled-channel
system. Moreover, in ref.~\cite{oset} it has been argued that, using unitarized ChPT (UChPT)
in a finite volume, it is possible first to directly fit the parameters of the chiral
potential to the energy spectrum measured on the lattice and eventually to determine
the physical observables  (the phase shifts, resonance parameters, etc) from the solution of
the scattering equations. The feasibility of such a procedure was demonstrated in the
example with synthetic data~\cite{oset}. 
For the further applications of this
method, see refs.~\cite{Torres:2011pr,kappa}.

Inelastic resonances, which have a significant decay rate into the 
three and more particle final states, 
have received much less attention in the literature so far.
The Roper resonance $N(1440)$, which
decays into the inelastic three-particle channels with an approximately
 40\,\% probability,
is an example of such a system.
{\it A priori,} one may expect significant finite-volume effects in this decay,
which can not be evaluated by using the standard L\"uscher approach. 
For this reason, at the moment it is not clear, whether the reverse level ordering
between $N(1440)$ and $N(1535)$ in lattice simulations, mentioned above,
can not be at least partially attributed to these finite-volume effects.
Putting it differently, one may ask, whether the effect still persists for the 
true resonance
positions in the infinite volume. We therefore conclude that it is highly
desirable to construct a framework that will allow one to systematically calculate the
finite-volume effects, coming from the tree-body final states.

Formulating a counterpart of the L\"uscher approach in a three-body case represents
a major challenge. For this reason, at the first step, we want to simplify the problem
as much as possible. Namely, we 
consider a non-relativistic quantum-mechanical model
with coupled two-particle and three-particle channels. 
Multi-particle channels (4 and more
particles) are neglected from the beginning. Using these approximations, 
 in particular,
the technical complications, related to the necessity of Lorentz-boosting 
the two-particle sub-systems in the three-particle state to their respective center-of-mass 
(CM) frames can be avoided. 
Moreover, using the potential model instead of the effective field theory (EFT)
framework, we avoid the discussion of the proper counting rules 
for the multi-particle
intermediate states, which are generated by loops. The core of the problem, which
consists in the study of the finite-volume effects coming from the three-particle
intermediate states, remains however unaffected by the above approximations. 
At the next step, we plan to carry out a  full-fledged investigation 
of the problem.

The central issue addressed in our study, can be formulated as follows. 
In case of
the elastic two-body scattering, the L\"uscher formula relates the discrete spectrum
in a finite box to the scattering $S$-matrix element in the infinite volume, 
up to the corrections that are exponentially suppressed in the 
box size. 
Putting it differently, this is a relation between the {\em observables}
in a finite and in the infinite volumes: the details of the potential do not matter
if the box size is much larger than the typical radius of interaction. It is intuitively
clear that the statement must remain valid even if there is a
contribution from the  three-particle intermediate state -- as far as the size
of the box remains much larger than the scales characterizing the interactions.
In our paper, we prove this statement for the model described in the previous
paragraph. 
It is expected that this proof can be extended beyond this particular
model.

The layout of the present paper is as follows. In section~\ref{sec:model} we discuss
the quantum-mechanical model which describes the scattering in the coupled two- and three-
particle channels. Considering the same model in a finite volume, we derive an equation
which predicts the volume-dependent energy levels. Further, in section~\ref{sec:twobody}
we re-derive the L\"uscher formula for the two-body scattering applying a new method,
which can be used with minor modifications in the 3-body case as well.
This is done in section~\ref{sec:threebody}, in which we derive the three-body analog
of the L\"uscher formula. Section~\ref{sec:concl} contains our conclusions.

\section{Three-body problem in a finite volume}
\label{sec:model}

\subsection{The model in the infinite volume}

In this section, we describe a model which will be later used to study 
the finite-volume effects in the three-body sector. 
In order to make the presentation self-contained, below we explicitly
display the standard formulas in case of the infinite volume.
We consider a non-relativistic quantum-mechanical system 
of three spinless non-identical particles with
the masses \textit{ $m_\alpha,~\alpha=1,2,3$}. The Hamiltonian of the model is given by a sum of a free Hamiltonian
\textit{ ${\bf H}_{\sf 0}$}, pair interaction Hamiltonian \textit{${\bf H}_{\sf 2\to 2}$} and the Hamiltonian 
\textit{${\bf H}_{\sf 2\to 3}$} that describes
the transition from two- to three-particle state \textit{$1+2\to 1+2+3$}.  
In order to simplify the following expressions,
we do not include the three-particle force into the Hamiltonian.
As shown later, the induced three-particle force will anyway emerge
from the two-particle interactions, so one could have indeed added it from the
beginning without much effort.
The explicit expression of the full Hamiltonian in terms
of the creation/annihilation operators is given by
\textit{\eq\label{eq:H}
{\bf H}&=&{\bf H}_{\sf 0}+{\bf H}_{\sf 2\to 2}+{\bf H}_{\sf 2\to 3}\doteq {\bf H}_{\sf 0}+{\bf H}_{\sf I}\, ,
\nonumber\\[2mm]
{\bf H}_{\sf 0}&=&\sum_{\alpha=1}^3\sum_{\bf k}
\biggl(m_\alpha+\dfrac{{\bf k}^2}{2m_\alpha}\biggr)\,a_\alpha({\bf k})a_\alpha^\dagger({\bf k})\, ,
\nonumber\\[2mm]
{\bf H}_{\sf 2\to 2}&=&\sum_{(\alpha\beta)=(12),(23),(13)}^3\sum_{{\bf k}_1\cdots {\bf k}_4}\,
(2\pi)^3\delta^3({\bf k}_1+{\bf k}_2-{\bf k}_3-{\bf k}_4)\,
V_{\alpha\beta}({\bf k}_1,{\bf k}_2;{\bf k}_3,{\bf k}_4)\,
\nonumber\\[2mm]
&\times&a_\alpha({\bf k}_1)a_\beta({\bf k}_2)a_\alpha^\dagger({\bf k}_3)a_\beta^\dagger({\bf k}_4)\, ,
\nonumber\\[2mm]
{\bf H}_{\sf 2\to 3}&=&\sum_{{\bf k}_1\cdots {\bf k}_5}\,
(2\pi)^3\delta^3({\bf k}_1+{\bf k}_2-{\bf k}_3-{\bf k}_4-{\bf k}_5)
\nonumber\\[2mm]
&\times&\biggl\{
\Gamma({\bf k}_1,{\bf k}_2;{\bf k}_3,{\bf k}_4,{\bf k}_5)\,
a_1({\bf k}_1)a_2({\bf k}_2)a_1^\dagger({\bf k}_3)a_2^\dagger({\bf k}_4)a_3^\dagger({\bf k}_5)
+\sf{h.c.}\biggr\}\, ,
\en }
where \textit{}$a_\alpha({\bf k})$ and  \textit{$a_\alpha^\dagger({\bf k})$} denote the annihilation and creation operators
for the particle $\alpha$, respectively. Further, \textit{$V_{12}\doteq V_3,~V_{23}\doteq V_1,~V_{31}=V_2$} denote the pair
potentials, so that 
\textit{${\bf H}_{\sf 2\to 2}={\bf V}_1+{\bf V}_2+{\bf V}_3=\sum_{\alpha=1}^3{\bf V}_\alpha$}.
Finally, the vertex \textit{$\Gamma$} describes the transition \textit{$1+2\to 1+2+3$}.

Next, we introduce the free and full resolvents, which are defined by
\textit{\eq
{\bf G}_{\sf 0}(z)=\dfrac{1}{{\bf H}_{\sf 0}-z-i0}\, ,\quad\quad
{\bf G}(z)=\dfrac{1}{{\bf H}-z-i0}\, .
\en }
The scattering matrix, which is expressed through the resolvents as
\textit{\eq
{\bf G}(z)={\bf G}_{\sf 0}(z)+{\bf G}_{\sf 0}(z){\bf T}(z){\bf G}_{\sf 0}(z)\, ,
\en }
obeys the Lippmann-Schwinger (LS) equation\footnote{Note that the sign convention in the LS equation
below coincides with the one in the field theory and is opposite to the one usually adopted in
the potential scattering theory.}
\textit{\eq
{\bf T}(z)=(-{\bf H}_{\sf I})+(-{\bf H}_{\sf I}){\bf G}_{\sf 0}(z){\bf T}(z)\, .
\en }
In this paper, we are primarily aimed at the modeling of the $\pi N$ scattering in the Roper 
resonance region -- that is, in the presence of an open three-particle channel.
To this end, we
consider the scattering amplitude for the process \textit{$1+2\to 1+2$}. Introducing
the projection operators \textit{${\bf P}$} and \textit{${\bf Q}=\mathbb{I}-{\bf P}$} (where \textit{$\mathbb{I}$} stands 
for the unit operator), which project
onto the two-particle state \textit{$|12\rangle$} and the three-particle state \textit{$|123\rangle$},
respectively, we obtain the equation for the effective two-body operator
\textit{${\bf T}_{\sf P}(z)={\bf P}{\bf T}(z){\bf P}$}
\textit{\eq
{\bf T}_{\sf P}(z)={\bf W}(z)+{\bf W}(z){\bf G}_{\sf P}(z){\bf T}_{\sf P}(z)\, ,\quad\quad
 {\bf W}(z)=(-{\bf H}_{\sf 2\to 2})+{\bf H}_{\sf 2\to 3}{\bf G}_3(z)({\bf H}_{\sf 2\to 3})^\dagger\, ,
\en }
where the three-particle Green's function obeys the equation
\textit{\eq
{\bf G}_3(z)={\bf G}_{\sf Q}+{\bf G}_{\sf Q}(-{\bf H}_{\sf 2\to 2}){\bf G}_3(z)\, ,
\en }
and the following notations are used
\textit{\eq
{\bf G}_{\sf P}={\bf P}{\bf G}_{\sf 0}(z){\bf P}\, ,\quad\quad
{\bf G}_{\sf Q}={\bf Q}{\bf G}_{\sf 0}(z){\bf Q}\, .
\en }
Note that the effective one-channel potential \textit{${\bf W}(z)$} becomes non-Hermitian
above the three-particle threshold \textit{$z>m_1+m_2+m_3$}.

Next, we define the three-particle scattering amplitude \textit{${\bf R}(z)$} through
\textit{\eq
{\bf G}_3(z)={\bf G}_{\sf Q}+{\bf G}_{\sf Q}{\bf R}(z){\bf G}_{\sf Q}\, .
\en }
The quantity \textit{${\bf R}(z)$} can be expressed as
\textit{\eq
{\bf R}(z)=\sum_{\alpha,\beta=1}^3{\bf M}_{\alpha\beta}(z)\, ,
\en }
where \textit{${\bf M}_{\alpha\beta}(z)$} obeys Faddeev equations (see, \textit{e.g.},~\cite{Belyaev})
\textit{\eq
{\bf M}_{\alpha\beta}(z)=\delta_{\alpha\beta}{\bf T}_\alpha(z)
+{\bf T}_\alpha(z){\bf G}_{\sf Q}(z)\sum_{\gamma=1}^3(1-\delta_{\alpha\gamma}){\bf M}_{\gamma\beta}(z)\, ,
\en }
and \textit{${\bf T}_\alpha(z)$} denote the scattering amplitudes in the channel \textit{$\alpha$}
\textit{\eq
{\bf T}_\alpha(z)=(-{\bf V}_\alpha)+(-{\bf V}_\alpha){\bf G}_{\sf Q}(z){\bf T}_\alpha(z)\, .
\en }
In order to write down these equations explicitly in momentum space, 
it is useful to work in Jacobi basis
\textit{\eq\label{eq:Jacobi}
{\bf P}={\bf k}_1+{\bf k}_2+{\bf k}_3\, ,
\quad
{\bf p}_\alpha={\bf k}_\alpha\, ,
\quad
{\bf q}_\alpha=\dfrac{m_\gamma{\bf k}_\beta-m_\beta{\bf k}_\gamma}{m_\gamma+m_\beta}\, ,
\quad\quad
(\alpha\beta\gamma)=(123),(231),(312)\, .
\en }
Separating the center-of-mass (CM) motion from the matrix elements
\textit{\eq
\langle {\bf k}_1' {\bf k}_2' {\bf k}_3'|{\bf M}_{\alpha\beta}(z)| {\bf k}_1 {\bf k}_2 {\bf k}_3\rangle
&=&(2\pi)^3\delta^3({\bf k}_1'+{\bf k}_2'+{\bf k}_3'-{\bf k}_1-{\bf k}_2-{\bf k}_3)
\langle {\bf p}_\alpha'{\bf q}_\alpha'|{\bf m}_{\alpha\beta}(z)|{\bf p}_\beta{\bf q}_\beta\rangle\, ,
\nonumber\\[2mm]
\langle {\bf k}_1' {\bf k}_2' {\bf k}_3'|{\bf G}_{\sf Q}(z)| {\bf k}_1 {\bf k}_2 {\bf k}_3\rangle
&=&(2\pi)^3\delta^3({\bf k}_1'+{\bf k}_2'+{\bf k}_3'-{\bf k}_1-{\bf k}_2-{\bf k}_3)
\nonumber\\[2mm]
&\times&\dfrac{(2\pi)^3\delta^3({\bf p}_\alpha'-{\bf p}_\alpha)\,(2\pi)^3\delta^3({\bf q}_\alpha'-{\bf q}_\alpha)}
{M+\dfrac{{\bf p}_\alpha^2}{2M_\alpha}+\dfrac{{\bf q}_\alpha^2}{2\mu_\alpha}-z-i0}\, ,
\nonumber\\[2mm]
\langle {\bf k}_1' {\bf k}_2' {\bf k}_3'|{\bf T}_\alpha(z)| {\bf k}_1 {\bf k}_2 {\bf k}_3\rangle
&=&(2\pi)^3\delta^3({\bf k}_1'+{\bf k}_2'+{\bf k}_3'-{\bf k}_1-{\bf k}_2-{\bf k}_3)
(2\pi)^3\delta^3({\bf p}_\alpha'-{\bf p}_\alpha)
\nonumber\\[2mm]
&\times&
\langle {\bf q}_\alpha'|\btau_\alpha\biggl(z-m_\alpha-\dfrac{{\bf p}_\alpha^2}{2M_\alpha}\biggr)
|{\bf q}_\alpha\rangle\, ,
\en }
where \textit{$\btau_\alpha(z)$} is the two-body scattering amplitude, and
\textit{\eq
M=m_\alpha+m_\beta+m_\gamma\, ,\quad\quad
M_\alpha=\dfrac{m_\alpha(m_\beta+m_\gamma)}{m_\alpha+m_\beta+m_\gamma}\, ,\quad\quad
\mu_\alpha=\dfrac{m_\beta m_\gamma}{m_\beta+m_\gamma}\, ,
\en }
we finally obtain
\textit{\eq\label{eq:Fm}
&&\langle {\bf p}_\alpha'{\bf q}_\alpha'|{\bf m}_{\alpha\beta}(z)|{\bf p}_\beta{\bf q}_\beta\rangle
=\delta_{\alpha\beta}(2\pi)^3\delta^3({\bf p}_\alpha'-{\bf p}_\alpha)
\langle {\bf q}_\alpha'|\btau_\alpha\biggl(z-m_\alpha-\dfrac{{\bf p}_\alpha^2}{2M_\alpha}\biggr)
|{\bf q}_\alpha\rangle
\nonumber\\[2mm]
&+&\sum_{\gamma=1}^3(1-\delta_{\alpha\gamma})\int
\dfrac{d^3{\bf p}_\alpha''}{(2\pi)^3}\,\dfrac{d^3{\bf q}_\alpha''}{(2\pi)^3}\,
\dfrac{d^3{\bf p}_\gamma''}{(2\pi)^3}\,\dfrac{d^3{\bf q}_\gamma''}{(2\pi)^3}\,
(2\pi)^3\delta^3({\bf p}_\alpha'-{\bf p}_\alpha'')
\langle {\bf q}_\alpha'|\btau_\alpha\biggl(z-m_\alpha-\dfrac{({\bf p}_\alpha')^2}{2M_\alpha}\biggr)
|{\bf q}_\alpha''\rangle
\nonumber\\[2mm]
&\times&
\dfrac{(2\pi)^3\delta^3({\bf p}_\alpha''-{\bf p}_\alpha''({\bf p}_\gamma'',{\bf q}_\gamma''))
(2\pi)^3\delta^3({\bf q}_\alpha''-{\bf q}_\alpha''({\bf p}_\gamma'',{\bf q}_\gamma''))}
{M+\dfrac{({\bf p}_\alpha'')^2}{2M_\alpha}+\dfrac{({\bf q}_\alpha'')^2}{2\mu_\alpha}-z-i0}
\langle {\bf p}_\gamma''{\bf q}_\gamma''|{\bf m}_{\gamma\beta}(z)|{\bf p}_\beta{\bf q}_\beta\rangle\, .
\en }
The relations between the momenta \textit{${\bf p}_\alpha,{\bf p}_\beta,{\bf p}_\gamma$} and 
\textit{${\bf q}_\alpha,{\bf q}_\beta,{\bf q}_\gamma$} for different
channels are given by
\textit{\eq\label{eq:channelmomenta}
{\bf p}_\alpha=-\dfrac{m_\alpha}{m_\alpha+m_\beta}\,{\bf p}_\gamma
+{\bf q}_\gamma\, ,\quad\quad
{\bf q}_\alpha=-\dfrac{m_\beta (m_\alpha+m_\beta+m_\gamma)}{(m_\alpha+m_\beta)(m_\beta+m_\gamma)}\,{\bf p}_\gamma-\dfrac{m_\gamma}{m_\beta+m_\gamma}\,{\bf q}_\gamma\, ,
\nonumber\\[2mm]
{\bf p}_\alpha=-\dfrac{m_\alpha}{m_\alpha+m_\gamma}\,{\bf p}_\beta
-{\bf q}_\beta\, ,\quad\quad
{\bf q}_\alpha=\dfrac{m_\gamma (m_\alpha+m_\beta+m_\gamma)}
{(m_\alpha+m_\gamma)(m_\beta+m_\gamma)}\,{\bf p}_\beta
-\dfrac{m_\beta}{m_\beta+m_\gamma}\,{\bf q}_\beta\, .
\en } 
Here, \textit{$(\alpha\beta\gamma)=$ $(123)$, $(231)$, $(312)$}.

Further, one may express the effective one-channel potential through the solution of
Faddeev equations. Removing first the CM motion, one gets
\textit{\eq
\langle {\bf k}_1' {\bf k}_2' |{\bf W}(z)| {\bf k}_1 {\bf k}_2 \rangle
=(2\pi)^3\delta^3({\bf k}_1'+{\bf k}_2'-{\bf k}_1-{\bf k}_2)
\langle {\bf q}_3'|{\bf w}(z)|{\bf q}_3\rangle\, ,
\en }
with
\textit{\eq\label{eq:w}
\langle {\bf q}'|{\bf w}(z)|{\bf q}\rangle
&=&-\bar V_3( {\bf q}',{\bf q})+\sum_{\alpha,\beta=1}^3\int
\dfrac{d^3{\bf p}_\alpha''}{(2\pi)^3}\,\dfrac{d^3{\bf q}_\alpha''}{(2\pi)^3}\,
\dfrac{d^3{\bf p}_\beta'''}{(2\pi)^3}\,\dfrac{d^3{\bf q}_\beta'''}{(2\pi)^3}
\nonumber\\[2mm]
&\times&
\bar\Gamma_\alpha({\bf q}';{\bf p}_\alpha''{\bf q}_\alpha'')
\langle {\bf p}_\alpha''{\bf q}_\alpha''|{\bf g}_{3,\alpha\beta}(z)|{\bf p}_\beta'''{\bf q}_\beta'''\rangle
(\bar\Gamma_\beta({\bf q};{\bf p}_\beta'''{\bf q}_\beta'''))^*\, .
\en }
In the above expressions, the following notations are used
\textit{\eq
\bar\Gamma_\alpha({\bf q};{\bf p}_\alpha{\bf q}_\alpha)
&=&\Gamma\biggl({\bf q},-{\bf q};
{\bf k}_\alpha={\bf p}_\alpha,{\bf k}_\beta={\bf q}_\alpha-\dfrac{m_\beta}{m_\beta+m_\gamma}\,{\bf p}_\alpha,
{\bf k}_\gamma=-{\bf q}_\alpha-\dfrac{m_\gamma}{m_\beta+m_\gamma}\,{\bf p}_\alpha\biggr)\, ,
\nonumber\\[2mm]
\bar V_\alpha( {\bf q}',{\bf q})&=&V_\alpha({\bf q}',-{\bf q}';{\bf q},-{\bf q})\, ,
\en }
and
\textit{\eq\label{eq:g3}
&&\langle {\bf p}_\alpha''{\bf q}_\alpha''|{\bf g}_{3,\alpha\beta}(z)|{\bf p}_\beta'''{\bf q}_\beta'''\rangle
=\delta_{\alpha3}\delta_{\beta3}
\dfrac{(2\pi)^3\delta^3({\bf p}_3''-{\bf p}_3''')
(2\pi)^3\delta^3({\bf q}_3''-{\bf q}_3''')}
{M+\dfrac{({\bf p}_3'')^2}{2M_3}+\dfrac{({\bf q}_3'')^2}{2\mu_3}-z-i0}
\nonumber\\[2mm]
&+&\dfrac{1}{M+\dfrac{({\bf p}_\alpha'')^2}{2M_\alpha}+\dfrac{({\bf q}_\alpha'')^2}{2\mu_\alpha}-z-i0}\,
\langle {\bf p}_\alpha''{\bf q}_\alpha''|{\bf m}_{\alpha\beta}(z)|{\bf p}_\beta'''{\bf q}_\beta'''\rangle\,
\dfrac{1}{M+\dfrac{({\bf p}_\beta''')^2}{2M_\beta}+\dfrac{({\bf q}_\beta''')^2}{2\mu_\beta}-z-i0}
\en }
Finally, the LS equation for the two-body scattering matrix \textit{${\bf T}_{\sf P}(z)$}
after removing the CM motion
\textit{\eq
\langle {\bf k}_1' {\bf k}_2' |{\bf T}_{\sf P}(z)| {\bf k}_1 {\bf k}_2 \rangle
=(2\pi)^3\delta^3({\bf k}_1'+{\bf k}_2'-{\bf k}_1-{\bf k}_2)
\langle {\bf q}_3'|{\bf t}_{\sf P}(z)|{\bf q}_3\rangle\, ,
\en }
takes the form
\textit{\eq\label{eq:LS-eff}
\langle {\bf q}'|{\bf t}_{\sf P}(z)|{\bf q}\rangle=\langle {\bf q}'|{\bf w}(z)|{\bf q}\rangle
+\int\dfrac{d^3{\bf q}''}{(2\pi)^3}\,
\dfrac{\langle {\bf q}'|{\bf w}(z)|{\bf q}''\rangle\,\langle {\bf q}''|{\bf t}_{\sf P}(z)|{\bf q}\rangle}
{m_1+m_2+\dfrac{({\bf q}'')^2}{2\mu_3}-z-i0}\, .
\en }
The above formulas simplify considerably, if the pair potentials have separable 
form. For completeness, in Appendix~\ref{app:separable} we list the
pertinent expressions in the separable model for the case of three 
identical particles.

\subsection{The model in a finite volume}
\label{sec:model-finite}

Now, let us put the system 
described by the Hamiltonian in eq.~(\ref{eq:H}), 
in a finite cubic box of a size \textit{$L$}. 
Assuming periodic boundary conditions,
the momenta of all free particles
take discrete values \textit{${\bf k}_\alpha=2\pi{\bf n}_\alpha/L\,,~\alpha=1,2,3$} and 
\textit{${\bf n}_\alpha\in\mathbb{Z}^3$}.
The only difference between the infinite-volume and finite-volume cases
consists in replacing the momentum-space integrals in all scattering equations 
by the sums over the discrete momenta.

The finite-volume counterpart of the 
LS equation with the effective two-body potential in eq.~(\ref{eq:LS-eff})
is given by 
\textit{\eq\label{eq:LS-effL}
\langle {\bf q}'|{\bf t}_{\sf P}^L(z)|{\bf q}\rangle=\langle {\bf q}'|{\bf w}^L(z)|{\bf q}\rangle
+\dfrac{1}{L^3}\sum_{{\bf q}''}\,
\dfrac{\langle {\bf q}'|{\bf w}^L(z)|{\bf q}''\rangle\,\langle {\bf q}''|{\bf t}_{\sf P}^L(z)|{\bf q}\rangle}
{m_1+m_2+\dfrac{({\bf q}'')^2}{2\mu_3}-z}\, .
\en }
Here and below, we attach the superscript ``$L$'' to the quantities
defined in a finite volume.
Note also that in the CM frame, the relative three momentum in the intermediate state is given by
\textit{${\bf k}_1''=-{\bf k}_2''={\bf q}''$}. This means, that the summation momentum
\textit{${\bf q}''$} takes the
discrete values \textit{${\bf q}''=2\pi{\bf n}/L\,,~{\bf n}\in \mathbb{Z}^3$}.

The scattering amplitude \textit{${\bf t}_{\sf P}^L(z)$} has an infinite tower of poles, corresponding to the
discrete energy spectrum of a system in a finite box. It can be shown that the 
locations of these poles are
determined by the L\"uscher formula. Indeed, 
performing the partial-wave expansion in eq.~(\ref{eq:LS-effL})
\textit{\eq
\langle {\bf q}'|{\bf t}_{\sf P}^L(z)|{\bf q}\rangle&=&4\pi\sum_{l'm',lm}
Y_{l'm'}(\hat{\bf q}')t^L_{l'm',lm}(q',q;z)Y_{lm}^*(\hat{\bf q})\, ,
\nonumber\\[2mm]
\langle {\bf q}'|{\bf w}^L(z)|{\bf q}\rangle&=&4\pi\sum_{l'm',lm}
Y_{l'm'}(\hat{\bf q}')w^L_{l'm',lm}(q',q;z)Y_{lm}^*(\hat{\bf q})\, ,
\en }
where \textit{$\hat {\bf q}$} denotes the unit vector in the direction of \textit{${\bf q}$} and 
\textit{$Y_{lm}(\hat {\bf q})$} stands for the spherical function.
The position of the poles is determined by the equation
\textit{\eq\label{eq:DM}
\det{\cal D}&=&0\, ,
\nonumber\\[2mm]
{\cal D}_{l'm',lm}&=&\delta_{l'l}\delta_{m'm}-\sum_{l''m''}\dfrac{\mu_3p}{2\pi}K^L_{l'm',l''m''}(p,p;z(p))
{\cal M}_{l''m'',lm}(\nu)\, ,
\nonumber\\[2mm]
{\cal M}_{l'm',lm}(\nu)&=&\dfrac{(-)^{l'}}{\pi^{3/2}}\,\sum_{j=|l'-l|}^{l'+l}\sum_{s=-j}^j\dfrac{i^j}{\nu^{j+1}}
Z_{js}(1;\nu^2)C_{l'm',js,lm}\, .
\en }
Here,
\textit{\eq
\nu=\dfrac{pL}{2\pi}\, ,
\en }
the quantity \textit{$Z_{js}(1;\nu^2)$} denotes the L\"uscher zeta-function,
and the symbols \textit{$C_{l'm',js,lm}$} are given by (see, \textit{e.g.}~\cite{luescher-torus})
\textit{\eq
C_{l'm',js,lm}=(-)^mi^{l'-j+l}\sqrt{(2l'+1)(2j+1)(2l+1)}
\begin{pmatrix}l' & j & l \\ m' & s & -m\end{pmatrix} 
\begin{pmatrix}l' & j & l \\ 0 & 0 & 0\end{pmatrix}\, . 
\en }
Further, the quantity \textit{$K^L(q,q,z(q))$} with \textit{$z(q)=m_1+m_2+\dfrac{q^2}{2\mu_3}$}
is the on-shell solution of the LS equation up to the exponentially suppressed contributions in the box size \textit{$L$}. It is given by
\textit{\eq\label{eq:KpartialL}
K^L_{l'm',lm}(q',q;z(q))&=&w^L_{l'm',lm}(q',q;z(q))+\dfrac{\mu_3}{2\pi}\,{\sf P.V.}\int_0^\infty\dfrac{(q'')^2dq''}{(q'')^2-q^2}\,
\nonumber\\[2mm]
&\times&\sum_{l''m''}w^L_{l'm',l''m''}(q',q'';z(q))K^L_{l''m'',lm}(q'',q;z(q))\, .
\en }
Here, \textit{${\sf P.V.}$} stands for the principal-value integral. 

In the above equations, the mixing of the partial waves occurs, because
the rotational symmetry is broken on the cubic lattice. The equation can
be partially diagonalized in the basis of the irreducible representations of the
cubic group~\cite{luescher-torus,Luu}. In case of the fermions, the basis of
the irreducible representations of the double cover of the cubic group
should be used~\cite{Lage-distributions}.

The expressions simplify considerably, if, \textit{e.g.}, one assumes that only the S-wave
scattering contributes. Then, the L\"uscher formula can be rewritten as
\textit{\eq
\phi(\nu)=-\delta^L(p)+\pi n\, ,\quad n=0,1,\cdots\, ,\quad
\tan\phi(\nu)=-\dfrac{\pi^{3/2}\nu}{Z_{00}(1;\nu^2)}\, .
\en }
The quantity \textit{$\delta^L(p)$} stands for the so-called {\em pseudophase}~\cite{lage-KN,scalar,oset}
\textit{\eq
\tan\delta^L(p)=\dfrac{\mu_3p}{2\pi}\,K^L_{00,00}(p,p;z(p))\, .
\en }
If the energy \textit{$z$} is below the three-particle threshold, the volume-dependence
in the effective two-body potential
is exponentially suppressed in \textit{$L$}. Consequently, up to the exponentially 
suppressed contributions, the pseudophase  
 \textit{$\delta^L(p)$} does not depend on \textit{$L$} and on the level index \textit{$n$} (the latter
dependence arises through the dependence of the effective potential
on \textit{$L=L_n(p)$} at a given value of \textit{$p$}). In this case,  \textit{$\delta^L(p)=\delta(p)$}
coincides with the conventional elastic scattering phase. 

Above the 3-body threshold, the effective potential is given by (cf. eq.~(\ref{eq:w}))
\textit{\eq\label{eq:wL}
\langle {\bf q}'|{\bf w}^L(z)|{\bf q}\rangle
&=&-\bar V_3( {\bf q}',{\bf q})+\dfrac{1}{L^{12}}\,\sum_{\alpha,\beta=1}^3
\sum_{{\bf p}_\alpha''{\bf q}_\alpha''{\bf p}_\beta'''{\bf q}_\beta'''}
\bar\Gamma_\alpha({\bf q}';{\bf p}_\alpha''{\bf q}_\alpha'')
\nonumber\\[2mm]
&\times&\langle {\bf p}_\alpha''{\bf q}_\alpha''|{\bf g}^L_{3,\alpha\beta}(z)|{\bf p}_\beta'''{\bf q}_\beta'''\rangle
(\bar\Gamma_\beta({\bf q};{\bf p}_\beta'''{\bf q}_\beta'''))^*\, .
\en }
Note that in the above equation, we do not attach the superscript \textit{$L$} to the quantities
\textit{$\bar V_3$} and \textit{$\bar\Gamma$}, which are the same in a finite and in the infinite
volumes. Further, the summation in eq.~(\ref{eq:wL}) runs over the momenta 
(cf. eq.~(\ref{eq:Jacobi}))
\textit{\eq\label{eq:JacobiL}
&&{\bf p}_\alpha''=\dfrac{2\pi{\bf n}_\alpha''}{L}\, ,\quad\quad
{\bf p}_\beta'''=\dfrac{2\pi{\bf n}_\beta'''}{L}\, ,
\nonumber\\[2mm]
&&{\bf q}_\alpha''=\dfrac{2\pi}{L}\,\biggl({\bf l}_\alpha''+\dfrac{m_\beta}{m_\beta+m_\gamma}{\bf n}_\alpha''\biggr)\, ,
\quad\quad
{\bf q}_\beta'''=\dfrac{2\pi}{L}\,\biggl({\bf l}_\beta'''+\dfrac{m_\gamma}{m_\gamma+m_\alpha}{\bf n}_\beta'''\biggr)\, ,
\nonumber\\[2mm]
&&{\bf n}_\alpha'',{\bf n}_\beta''',{\bf l}_\alpha'',{\bf l}_\beta'''\in\mathbb{Z}^3\, .
\en }
The finite-volume version of eq.~(\ref{eq:g3}) reads
\textit{\eq\label{eq:g3L}
&&\langle {\bf p}_\alpha''{\bf q}_\alpha''|{\bf g}^L_{3,\alpha\beta}(z)|{\bf p}_\beta'''{\bf q}_\beta'''\rangle
=\dfrac{\delta_{\alpha3}\delta_{\beta3}
L^3\delta_{{\bf p}_3''{\bf p}_3'''}
L^3\delta_{{\bf q}_3''{\bf q}_3'''}}
{M+\dfrac{({\bf p}_3'')^2}{2M_3}+\dfrac{({\bf q}_3'')^2}{2\mu_3}-z}
\nonumber\\[2mm]
&+&\dfrac{1}{M+\dfrac{({\bf p}_\alpha'')^2}{2M_\alpha}+\dfrac{({\bf q}_\alpha'')^2}{2\mu_\alpha}-z}\,
\langle {\bf p}_\alpha''{\bf q}_\alpha''|{\bf m}^L_{\alpha\beta}(z)|{\bf p}_\beta'''{\bf q}_\beta'''\rangle\,
\dfrac{1}{M+\dfrac{({\bf p}_\beta''')^2}{2M_\beta}+\dfrac{({\bf q}_\beta''')^2}{2\mu_\beta}-z}\, .
\en }
Faddeev equations in a finite volume are written as (cf. eq.~(\ref{eq:Fm}))
\textit{\eq\label{eq:FmL}
\langle {\bf p}_\alpha'{\bf q}_\alpha'|{\bf m}^L_{\alpha\beta}(z)|{\bf p}_\beta{\bf q}_\beta\rangle
&=&\delta_{\alpha\beta}L^3\delta_{{\bf p}_\alpha'{\bf p}_\alpha}
\langle {\bf q}_\alpha'|\btau^L_\alpha(z;{\bf p}_\alpha)|{\bf q}_\alpha\rangle
+
\dfrac{1}{L^{12}}\sum_{\gamma=1}^3(1-\delta_{\alpha\gamma})
\sum_{{\bf p}_\alpha'',{\bf q}_\alpha'',{\bf p}_\gamma'',{\bf q}_\gamma''}
L^3\delta_{{\bf p}_\alpha'{\bf p}_\alpha''}
\nonumber\\[2mm]
&&\hspace*{-2.5cm}\times\,\,\langle {\bf q}_\alpha'|\btau^L_\alpha(z;{\bf p}_\alpha)
|{\bf q}_\alpha''\rangle\,
\dfrac{L^3\delta_{{\bf p}_\alpha'',{\bf p}_\alpha''({\bf p}_\gamma'',{\bf q}_\gamma'')}
L^3\delta_{{\bf q}_\alpha'',{\bf q}_\alpha''({\bf p}_\gamma'',{\bf q}_\gamma'')}}
{M+\dfrac{({\bf p}_\alpha'')^2}{2M_\alpha}+\dfrac{({\bf q}_\alpha'')^2}{2\mu_\alpha}-z}
\langle {\bf p}_\gamma''{\bf q}_\gamma''|{\bf m}^L_{\gamma\beta}(z)|{\bf p}_\beta{\bf q}_\beta\rangle\, ,
\en }
where the two-body scattering amplitude in a finite volume is a solution of the equation
\textit{\eq\label{eq:twobodyL}
&& \langle {\bf q}_\alpha'|\btau^L_\alpha(z;{\bf p}_\alpha)|{\bf q}_\alpha\rangle
=(-\bar V_\alpha ({\bf q}_\alpha',{\bf q}_\alpha))
+
\dfrac{1}{L^3}\,\sum_{{\bf q}_\alpha''}\dfrac{(-\bar V_\alpha ({\bf q}_\alpha',{\bf q}_\alpha''))
 \langle {\bf q}_\alpha''|\btau^L_\alpha(z;{\bf p}_\alpha)|{\bf q}_\alpha\rangle}
{M+\dfrac{{\bf p}_\alpha^2}{2M_\alpha}+\dfrac{({\bf q}_\alpha'')^2}{2\mu_\alpha}-z}\, .
\en }
Note that the momentum \textit{${\bf p}_\alpha$} enters in the definition of \textit{$\btau^L_\alpha$}
twice: in the argument \textit{$z\to z-m_\alpha-\dfrac{{\bf p}_\alpha^2}{2M_\alpha}$}, and
through the shifting of the summation over the momenta \textit{${\bf q}_\alpha''$} which, according
to eq.~(\ref{eq:JacobiL}), is no longer proportional to an integer number. In the infinite
volume, the summation is replaced by an integration over the whole momentum space,
and the dependence on  \textit{${\bf p}_\alpha$} remains only in the argument.

As in the infinite volume, the equations simplify considerably, if we assume
pair interactions of the separable form. The pertinent equations are 
listed in Appendix~\ref{app:separableL}.

\subsection{Singularities of the effective two-body potential}

As mentioned above, below the three-particle threshold the finite-volume
contributions to \textit{${\bf w}^L(z)$} are exponentially suppressed. Therefore,
up to such suppressed terms, the effective potential coincides with the
regular function \textit{${\bf w}(z)$}, which is defined in the infinite volume.

Above the three-particle threshold, the singularities emerge in 
\textit{${\bf w}^L(z)$}. In the vicinity of these singularities the pseudophase
rapidly changes by \textit{$\pi$}. These singularities may strongly affect the
finite volume spectrum in the vicinity and above the inelastic threshold.
For this reason, below we shall discuss them in detail.

Potentially, the matrix element \textit{$\langle {\bf q}'|{\bf w}^L(z)|{\bf q}\rangle$}
may become singular, when

\begin{itemize}

\item[i)]
the three-particle denominators \textit{$M+\dfrac{{\bf p}_\alpha^2}{2M_\alpha}+
\dfrac{{\bf q}_\alpha^2}{2\mu_\alpha}-z$} in eq.~(\ref{eq:g3L}) vanish;

\item[ii)]
the matrix elements of the operator \textit{${\bf m}^L_{\alpha\beta}(z)$}, which are
solutions of the Faddeev equations in a finite volume, develop a pole at 
those values of energy, which do not coincide with the poles of the 
three-particle energy denominators.

\end{itemize}

We are going to demonstrate below that the singularities of the first type
cancel, and only of the second type survive. In order to prove this statement,
we mention that, from eqs.~(\ref{eq:JacobiL}) and (\ref{eq:twobodyL}) it
follows that the two-body scattering matrix \textit{$\btau_\alpha^L(z)$} vanishes exactly
for those values of \textit{$z$} where the three-particle propagators develop a pole.
These energies are given by the following equation
\textit{\eq\label{eq:cancel}
z=M+\dfrac{({\bf p}_\alpha'')^2}{2M_\alpha}+\dfrac{1}{2\mu_\alpha}\,
\biggl({\bf p}_\gamma''+\dfrac{m_\gamma}{m_\beta+m\gamma}\,
{\bf  p}_\alpha''\biggr)^2\, ,\quad
 ({\bf p}_\alpha'',{\bf p}_\gamma'')
=\dfrac{2\pi}{L}\,({\bf n}_\alpha'',{\bf n}_\gamma'')\, ,\quad
{\bf n}_\alpha'',{\bf n}_\gamma''\in\mathbb{Z}^3\, .
\en }
Let us show now that there are no three-particle singularities of the first type
in the matrix elements of the three-particle Green's function. 
For simplicity, we check this property only for the lowest
three-body singularity at \textit{$z=M$}.

\begin{figure}[t]
\begin{center}
\includegraphics[width=16.cm]{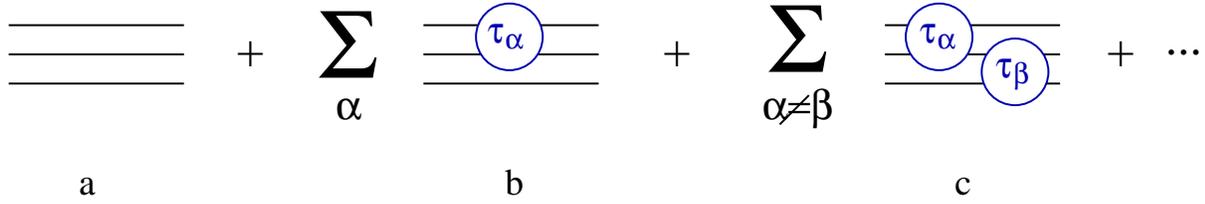}
\end{center}
\caption{A multiple-scattering series for the three-body Green's function.}
\label{fig:g3Lz}
\end{figure}

Let us consider the multiple scattering series for the three-body Green's
function
which is shown in Fig.~\ref{fig:g3Lz}. According to this, the sum of the matrix
elements of  \textit{${\bf g}_{3\alpha\beta}^L(z)$} is written as
(cf. eq.~(\ref{eq:g3L}))
\textit{\eq\label{eq:multipleL}
&&\sum_{\alpha,\beta=1}^{3}\langle {\bf p}_\alpha'{\bf q}_\alpha'|
{\bf g}_{3\alpha\beta}^L(z)|{\bf p}_\beta{\bf q}_\beta\rangle
=\dfrac{L^3\delta_{{\bf p}_3'{\bf p}_3}L^3\delta_{{\bf q}_3'{\bf q}_3}}
{M+\dfrac{{\bf p}_3^2}{2M_3}+\dfrac{{\bf q}_3^2}{2\mu_3}-z}
\nonumber\\[2mm]
&+&\sum_{\alpha=1}^3\dfrac{
L^3\delta_{{\bf p}_\alpha'{\bf p}_\alpha}
\langle {\bf q}_\alpha'|\btau^L_\alpha(z;{\bf p}_\alpha)
|{\bf q}_\alpha\rangle}{\biggl(M+\dfrac{({\bf p}_3')^2}{2M_3}+\dfrac{({\bf
      q}_3')^2}{2\mu_3}-z\biggr)\biggl(M+\dfrac{{\bf p}_3^2}{2M_3}+\dfrac{{\bf
      q}_3^2}{2\mu_3}-z\biggr)}+\cdots\, .
\en }
Here, we have used the property
\textit{\eq
M+\dfrac{{\bf p}_\alpha^2}{2M_\alpha}+\dfrac{{\bf q}_\alpha^2}{2\mu_\alpha}
=M+\dfrac{{\bf p}_\beta^2}{2M_\beta}+\dfrac{{\bf q}_\beta^2}{2\mu_\beta}\, ,
\quad\quad
\alpha,\beta=1,2,3\quad\mbox{and}\quad \alpha\neq\beta\, .
\en }
The first term in eq.~(\ref{eq:multipleL}) has a pole at \textit{$z=M$}, when
\textit{${\bf p}_3'={\bf p}_3={\bf q}_3'={\bf q}_3={\bf 0}$}. The denominator
in the second term has a double zero here, but the matrix element
of the operator \textit{$\btau^L_\alpha(z;{\bf p}_\alpha)$} also has a zero, so that only a single
pole remains. This property can be easily checked in higher orders.
Introducing the notation
\textit{\eq
y_\alpha=\dfrac{1}{L^3}\,\dfrac{\langle {\bf 0}|\btau^L_\alpha(z;{\bf 0})|{\bf 0}\rangle}
{M-z}\, ,
\en }
we obtain (see eqs.~(\ref{eq:g3L}) and (\ref{eq:FmL}))
\textit{\eq\label{eq:appendix}
\sum_{\alpha,\beta=1}^{3}\langle {\bf 0}{\bf 0}|
{\bf g}_{3\alpha\beta}^L(z)|{\bf 0}{\bf 0}\rangle
&=&\dfrac{L^6}{M-z}\,\biggl\{1+(y_1+y_2+y_3)
\nonumber\\[2mm]
&+&(y_1(y_2+y_3)+y_2(y_1+y_3)+y_3(y_1+y_2))+\cdots\biggr\}+\mbox{\sf non-sing}\,
,
\en }
where ``non-sing'' stands for the terms that are non-singular at \textit{$z=M$}.

Summing up the series  
(see Appendix~\ref{app:summation}), we find
\textit{\eq\label{eq:summation}
\sum_{\alpha,\beta=1}^{3}\langle {\bf 0}{\bf 0}|
{\bf g}_{3\alpha\beta}^L(z)|{\bf 0}{\bf 0}\rangle
=\dfrac{L^6}{M-z}\,\dfrac{x_1x_2x_3}{x_1x_2+x_2x_3+x_3x_1-2x_1x_2x_3}+\mbox{\sf non-sing}\, ,
\en }
where \textit{$x_\alpha=1+y_\alpha$}. 
Let us now show that, in fact, the first term in eq.~(\ref{eq:summation})
is also non-singular. To this end, we consider the equation for the operator
\textit{$\btau^L_{\alpha}$}
\textit{\eq\label{eq:twobodyLx}
&& \langle {\bf q}_\alpha'|\btau^L_{\alpha}(z;{\bf 0})|{\bf q}_\alpha\rangle
=(-\bar V_\alpha ({\bf q}_\alpha',{\bf q}_\alpha))
+
\dfrac{1}{L^3}\,\sum_{{\bf q}_\alpha''}\dfrac{(-\bar V_\alpha ({\bf q}_\alpha',{\bf q}_\alpha''))
 \langle {\bf q}_\alpha''|\btau^L_{\alpha}(z;{\bf 0})|{\bf q}_\alpha\rangle}
{M+\dfrac{({\bf q}_\alpha'')^2}{2\mu_\alpha}-z}\, .
\en }
In order to separate the singularity at \textit{$z=M$}, we single out the term
\textit{${\bf q}''_\alpha=0$} in the above sum. It is then straightforward to see that
\textit{\eq\label{eq:yLR}
y_\alpha=\dfrac{y_{{\sf R}\alpha}}{1-y_{{\sf R}\alpha}}\, ,\quad\quad
y_{{\sf R}\alpha}=\dfrac{1}{L^3}\,
\dfrac{\langle {\bf 0}|\btau^L_{{\sf R}\alpha}(z;{\bf 0})|{\bf 0}\rangle}
{M-z}\, ,
\en }
where
\textit{\eq\label{eq:twobodyLR}
&& \langle {\bf q}_\alpha'|\btau^L_{{\sf R}\alpha}(z;{\bf 0})|{\bf q}_\alpha\rangle
=(-\bar V_\alpha ({\bf q}_\alpha',{\bf q}_\alpha))
+
\dfrac{1}{L^3}\,\sum_{{\bf q}_\alpha''\neq {\bf 0}}\dfrac{(-\bar V_\alpha ({\bf q}_\alpha',{\bf q}_\alpha''))
 \langle {\bf q}_\alpha''|\btau^L_{{\sf R}\alpha}(z;{\bf 0})|{\bf q}_\alpha\rangle}
{M+\dfrac{({\bf q}_\alpha'')^2}{2\mu_\alpha}-z}\, .
\en }
Since the term with \textit{${\bf q}_\alpha''= {\bf 0}$} is absent in eq.~(\ref{eq:twobodyLR}), the quantity 
\textit{$\langle {\bf q}_\alpha'|\btau^L_{{\sf R}\alpha}(z;{\bf 0})|{\bf q}_\alpha\rangle$}
is non-singular at \textit{$z\to M$}. Now, the validity of our statement after eq.~(\ref{eq:summation})
follows from the fact that \textit{$x_\alpha=1+y_\alpha=O(z-M)$} as \textit{$z\to M$}
and, consequently,
the three-particle singularities cancel in the sum of the matrix elements
of \textit{${\bf g}_{3\alpha\beta}^L(z)$}.

In conclusion of this section we summarize the main points:

\begin{itemize}

\item[i)]
The singularities of the
effective two-body potential \textit{${\bf w}^L(z)$} are caused by
the singularities of the matrix elements of the operator 
\textit{${\bf m}_{\alpha\beta}(z)$} which emerge as a result of solution of
Faddeev equations in a finite volume and thus have a non-perturbative origin.

\item[ii)]
In other words, the singularities do not emerge, when the energy coincides
with the eigenvalues of the {\em free} Hamiltonian in the box. The 
singularity structure is determined by the eigenvalues of the {\em full}
Hamiltonian. 

\item[iii)]
Individual terms in the finite-volume multiple-scattering series, however, 
contain the singularities determined by the {\em free} Hamiltonian. 
In order to make them disappear, multiple-scattering series should be summed up
to all orders. The L\"uscher formula ensures such a summation in the two-particle
case. In this section we have explicitly 
demonstrated that the summation can cure the problem in the three-particle case
as well.

\item[iv)]
The results of this section may serve as a warning against an 
approximate truncation of the multiple-scattering series in a finite volume.
While in the infinite volume this may affect only the numerical precision,
the singularity structure of the effective potential can be modified 
drastically as a result of such a truncation in a finite volume. 
Consequently, if the approximations are made in the three-body case, 
the singularity structure of the effective potential
should be examined very carefully.

\end{itemize}

\section{An alternative derivation of the L\"uscher formula}
\label{sec:twobody}

In order to find the energy spectrum in a finite volume,
the equations, which are considered in 
section~\ref{sec:model-finite}, can be solved numerically, applying, \textit{e.g.},
the method of ref.~\cite{Doring:2011ip}. 
However, these equations contain potentials as well as off-shell two-body
scattering matrices, which are model-dependent. The central question 
is, whether the predicted energy levels are also model-dependent.
In other words, if two different potential models lead to the same $S$-matrix
in the infinite volume, can the finite-volume spectra in these models
be different? 

In case of the two-particle elastic scattering, the answer is given by
the L\"uscher formula, which relates the finite-volume spectrum to the
(on-shell) \textit{$S$}-matrix element. Our aim is to rewrite the three-particle equations
in a finite volume in a similar fashion, in terms of the on-shell \textit{$S$}-matrix
elements only. In order to do this, in this section we consider a 
novel derivation
of the L\"uscher formula. The method used here can be generalized for the case
of three particles, as shown in section~\ref{sec:threebody}.

Let us consider the sum
\textit{\eq\label{eq:sum2}
S_2=\dfrac{1}{L^3}\sum_{{\bf p}}\dfrac{\Phi({\bf p})}{{\bf p}^2-q_0^2}\, ,
\quad\quad {\bf p}=\dfrac{2\pi{\bf n}}{L}\, ,\quad{\bf n}\in \mathbb{Z}^3\, ,
\en }
where \textit{$\Phi({\bf p})$} denotes a regular function of \textit{${\bf p}$}. Next, we perform
a partial-wave expansion
\textit{\eq
\Phi({\bf p})=\sum_{lm}\dfrac{\Phi_l(p)}{p^l}\,{\cal Y}_{lm}({\bf p})\, ,
\quad\quad p=|{\bf p}|,
\en }
where \textit{${\cal Y}_{lm}({\bf p})=p^lY_{lm}(\hat {\bf p})$}.
The sum in eq.~(\ref{eq:sum2}) can be rewritten in  
the following form
\textit{\eq\label{eq:sum2-1}
S_2&=&\dfrac{1}{L^3}\sum_{\bf p}^{|{\bf p}|<\Lambda}\sum_{lm}\dfrac{{\cal Y}_{lm}({\bf p})}
{{\bf p}^2-q_0^2}\,\biggl(\dfrac{\Phi_l(p)}{p^l}
-\dfrac{\Phi_l(q_0)}{q_0^l}f(q_0^2/\mu^2)\biggr)
\nonumber\\[2mm]
&+&\dfrac{\Phi_l(q_0)}{q_0^l}f(q_0^2/\mu^2)\,
\dfrac{1}{L^3}\sum_{\bf p}^{|{\bf p}|<\Lambda}\sum_{lm}\dfrac{{\cal Y}_{lm}({\bf p})}
{{\bf p}^2-q_0^2}\, .
\en }
Note that we have introduced the momentum cutoff \textit{$\Lambda$}, in order 
to regularize intermediate expressions. The cutoff disappears from the final expressions\footnote{At finite values of \textit{$\Lambda$}, rapidly oscillating 
terms at \textit{$L\to\infty$} may occur for the sharp cutoff, so a mathematically 
rigorous procedure is to use a smooth cutoff at a momentum scale \textit{$\Lambda$}. 
In order to ease the notations, we however proceed further with s sharp cutoff. 
The oscillating terms are briefly considered in Appendix~\ref{app:L-2}.
It is shown there that these -- as expected -- are harmless.}, 
if the function \textit{$\Phi({\bf p})$} falls off sufficiently fast
with \textit{${\bf p}$} Further, we have introduced a regulator \textit{$f(x)$} with the following
properties:
\begin{enumerate}
\item
The function \textit{$f(x)$} is bounded and smooth (has any number of derivatives)
on the whole interval \textit{$x\in\, ]-\infty,\infty[$}.
\item
\textit{$f(x)=1$} if \textit{$x\geq 0$}.
\item
\textit{$\lim_{x\to 0^-}f(x)=1$} and \textit{$\lim_{x\to 0^-}f^{(n)}(x)=0$} for all \textit{$n\neq 0$}.
\item
The function \textit{$f(x)$} vanishes exponentially when \textit{$x\to-\infty$}.
\end{enumerate}
Otherwise, the function \textit{$f(x)$} is arbitrary. An example of such a function is
\textit{\eq
f(x)=\left\{
\begin{array}{l l l}
1 &,\quad\mbox{if}\quad& x\geq 0\\
\exp\biggl(\dfrac{1}{1-\exp(x^{-2})}\biggr) &,\quad\mbox{if}\quad& x<0\, .
\end{array}
\right.
\en }
The choice of the scale \textit{$\mu$} in the regulator \textit{$f$} is also arbitrary. For example, one 
could choose \textit{$\mu$} to coincide with the mass of the lightest particle.

Taking now into account the fact
that the expression\textit{ $\Phi_l(p)/p^l$} is a regular function of \textit{$p^2$} in the 
vicinity of \textit{$p^2=0$}, the regular summation theorem~\cite{luescher-2.} can be applied.
Up to the exponentially suppressed terms in \textit{$L$}, one may replace the first sum
in eq.~(\ref{eq:sum2-1}) by the integral
\textit{\eq
S_2&=&\int^{|{\bf p}|<\Lambda}\dfrac{d^3{\bf p}}{(2\pi)^3}\,\sum_{lm}\dfrac{{\cal Y}_{lm}({\bf p})}
{{\bf p}^2-q_0^2-i0}\,\biggl(\dfrac{\Phi_l(p)}{p^l}-\dfrac{\Phi_l(q_0)}{q_0^l}
f(q_0^2/\mu^2)\biggr)
\nonumber\\[2mm]
&+&\dfrac{\Phi_l(q_0)}{q_0^l}f(q_0^2/\mu^2)\,
\dfrac{1}{L^3}\sum_{\bf p}^{|{\bf p}|<\Lambda}\sum_{lm}\dfrac{{\cal Y}_{lm}({\bf p})}
{{\bf p}^2-q_0^2}\, ,
\en }
where \textit{$q_0^2\to q_0^2+i0$} prescription has been chosen arbitrarily 
(the numerator of the integrand vanishes at \textit{$p^2=q_0^2$}, 
so the prescription does not matter). Simplifying the above expression, 
we arrive at
\textit{\eq\label{eq:S2}
S_2&=&\int\dfrac{d^3{\bf p}}{(2\pi)^3}\,\dfrac{\Phi({\bf p})}{{\bf p}^2-q_0^2-i0}
+\dfrac{\sqrt{-q_0^2-i0}}{(4\pi)^{3/2}}\,\Phi_0(q_0)f(q_0^2/\mu^2)
\nonumber\\[2mm]
&+&\dfrac{1}{4\pi^2L}f(q_0^2/\mu^2)\sum_{lm}\dfrac{\Phi_l(q_0)}{\nu^l}\,Z_{lm}(1;\nu^2)\, ,
\en }
where \textit{$\nu=q_0L/(2\pi)$}, and \textit{$Z_{lm}(1;\nu^2)$} stands for the L\"uscher
zeta-function
\textit{\eq\label{eq:zeta}
Z_{lm}(1;\nu^2)=\lim_{\lambda\to\infty}\biggl\{\sum_{{\bf n}\in \mathbb{Z}^3}
\theta(\lambda^2-{\bf n}^2)\,
\dfrac{{\cal Y}_{lm}({\bf n})}{{\bf n}^2-\nu^2}-\delta_{l0}\delta_{m0}
\sqrt{4\pi}\lambda\biggr\}\, ,\quad\quad 
\lambda=\dfrac{\Lambda L}{2\pi}\, .
\en }
Formally, the above equation can be rewritten in the following manner
\textit{\eq\label{eq:formally}
\dfrac{1}{L^3}\sum_{\bf k}\dfrac{(2\pi)^3\delta^3({\bf p}-{\bf k})}{{\bf k}^2-q_0^2}
&=&\mbox{\sf P.V.}\,\dfrac{1}{{\bf p}^2-q_0^2}
+\sum_{lm}\dfrac{2}{\nu^{l+1}}\,Y^*_{lm}(\hat {\bf p})
Z_{lm}(1;\nu^2)\Delta({\bf p}^2,q_0^2)\, ,
\nonumber\\[2mm]
\mbox{\sf P.V.}\,\dfrac{1}{{\bf p}^2-q_0^2}&=&
\dfrac{1}{{\bf p}^2-q_0^2-i0}-i\pi\Delta({\bf p}^2,q_0^2)\, ,
\en }
where the quantity \textit{$\Delta({\bf p}^2,q_0^2)$} coincides with the conventional
Dirac \textit{$\delta$}-function \textit{$\delta({\bf p}^2-q_0^2)$}
for \textit{$q_0^2\geq 0$} and is defined through the action on
the smooth test functions
\textit{\eq\label{eq:testfunction}
-i\pi\int\dfrac{d^3{\bf p}}{(2\pi)^3}\,\Delta({\bf p}^2,q_0^2)\Phi({\bf p})
=\dfrac{\sqrt{-q_0^2-i0}}{(4\pi)^{3/2}}\,\Phi_0(q_0)f(q_0^2/\mu^2)
\en }
(recall that \textit{$f(q_0^2/\mu^2)=1$} for \textit{$q_0^2\geq 0$}).
For \textit{$q_0^2<0$}, the action of the distribution \textit{$\Delta$} on a test function
is defined through the analytic continuation of \textit{$\Phi_0(q_0)$} in 
eq.~(\ref{eq:testfunction}). The regulator \textit{$f$} serves the purpose to effectively cut the contributions
with \textit{$-q_0^2>\mu^2$}.

To summarize,  the momentum-space 
two-body Green's function in a finite volume, which is given by
\textit{\eq\label{eq:twobody-def-2}
 G_0^L({\bf k};z)
=\dfrac{2\mu}{L^3}\,\sum_{\bf p}\dfrac{(2\pi)^3\delta^3({\bf p}-{\bf k})}{{\bf p}^2-q_0^2}\, ,
\en }
where \textit{$\mu$} is the reduced mass and \textit{$z=m_1+m_2+\dfrac{q_0^2}{2\mu}$}, can be
decomposed into the following parts
\textit{\eq\label{eq:split-2}
G_0^L({\bf k};z)&=&2\mu\biggl\{\dfrac{1}{{\bf k}^2-q_0^2-i0}-i\pi\Delta({\bf k}^2,q_0^2)
+\Delta({\bf k}^2,q_0^2)\sum_{lm}Y^*_{lm}(\hat{\bf k})\dfrac{2}{\nu^{l+1}}\,
Z_{lm}(1;\nu^2)\biggr\}
\nonumber\\[2mm]
&\doteq&G_0({\bf k};z)+G_{\sf U}({\bf k};z)+G_{\sf F}({\bf k};z)
\doteq G_{\sf K}({\bf k};z)+G_{\sf F}({\bf k};z)\, .
\en }
The equation~(\ref{eq:split-2}) gives the splitting of the finite-volume
two-body Green's function into the infinite-volume part \textit{$G_{\sf K}=G_0+G_{\sf U}$}
(corresponding to the principal-value prescription at the singularity)
and the correction \textit{$G_{\sf F}$}.
The latter is proportional to the on-shell factor
\textit{$\Delta({\bf k}^2,q_0^2)$}, which projects onto the energy shell.
This fact plays a crucial role in the proof
of the statement that the finite-volume spectrum is determined only
by the infinite-volume \textit{$S$}-matrix elements which are defined on shell. 

The L\"uscher formula can be derived straightforwardly, using
eq.~(\ref{eq:split-2}). To this end, we take \textit{$q_0^2\geq 0$}. Then, 
\textit{$\Delta({\bf k}^2,q_0^2)$} is replaced by the conventional \textit{$\delta$}-function.
 First, note that, substituting the free Green's function by the quantity
defined in 
eq.~(\ref{eq:twobody-def-2}), the two-body LS equation in a finite volume
can be written formally in the same form as in the infinite volume 
\textit{\eq
\langle {\bf p}|{\bf T}^L(z)|{\bf q}\rangle&=&
\langle {\bf p}|(-{\bf V})|{\bf q}\rangle+
\int\dfrac{d^3{\bf k}}{(2\pi)^3}\, \langle {\bf p}|(-{\bf V})|{\bf k}\rangle
G_0^L({\bf k};z)\langle {\bf k}|{\bf T}^L(z)|{\bf q}\rangle\, .
\en }
Next, we define
\textit{\eq\label{eq:TKTL}
{\bf T}&=&(-{\bf V})+(-{\bf V}){\bf G}_0{\bf T}\, ,
\nonumber\\[2mm]
{\bf K}&=&(-{\bf V})+(-{\bf V})({\bf G}_0+{\bf G}_{\sf U}){\bf K}={\bf T}+{\bf T}{\bf G}_{\sf U}{\bf K}\, ,
\nonumber\\[2mm]
{\bf T}^L&=&(-{\bf V})+(-{\bf V})({\bf G}_0+{\bf G}_{\sf U}+{\bf G}_{\sf F}){\bf T}^L={\bf K}+{\bf K}{\bf G}_{\sf F}{\bf T}^L\, .
\en }
It is immediately seen that \textit{${\bf T}$} and \textit{${\bf K}$} are the two-body \textit{$T$}- and
\textit{$K$}-matrices in the infinite volume, respectively. 
The relation between the \textit{$K$}-matrix in the infinite volume
and the $T$-matrix in a finite volume \textit{${\bf T}^L$} is given by the last equation
in eq.~(\ref{eq:TKTL}). After carrying out the partial-wave expansion
\textit{\eq 
\langle {\bf p}|{\bf T}^L(z)|{\bf q}\rangle
&=&4\pi\sum_{l'm'}\sum_{lm}Y_{l'm'}(\hat{\bf p})T^L_{l'm',lm}(p,q;z)
Y_{lm}^*(\hat {\bf q})\, ,
\nonumber\\[2mm]
\langle {\bf p}|{\bf K}(z)|{\bf q}\rangle
&=&4\pi\sum_{lm}Y_{lm}(\hat {\bf p})K_l(p,q;z)
Y_{lm}^*(\hat {\bf q})\, ,
\en }
and integrating over the angles, this equation can be rewritten 
in the algebraic form
\textit{\eq\label{eq:K-elastic}
T^L_{l'm',lm}(p,q;z)&=&\delta_{ll'}\delta_{mm'}K_l(p,q;z)+
\dfrac{\mu q_0}{2\pi}\, K_{l'}(p,q_0;z)\sum_{l''m''}{\cal M}_{l'm',l''m''}(\nu)
T^L_{l''m'',lm}(q_0,q,q_0^2)\, ,
\nonumber\\
&&
\en }
where the matrix \textit{${\cal M}_{l'm',lm}(\nu)$} is displayed in eq.~(\ref{eq:DM}).
Going to the mass shell \textit{$p=q=q_0$} and requiring that the above system of
linear equations is singular (\textit{$T^L$} becomes infinity, that means that
 the determinant of
the above system of linear equations vanishes), we finally arrive
at the L\"uscher formula, see eq.~(\ref{eq:DM}). Note that, since
in case of an elastic two-body scattering considered here, the \textit{$K$}-matrix 
in the L\"uscher formula stands for the infinite-volume \textit{$K$}-matrix,
\textit{$K_{l'm',lm}^L$} in eq.~(\ref{eq:DM}) should be
substituted by \textit{$\delta_{ll'}\delta_{mm'}K_l$}.
Note also that, within the 
normalization used, the on-shell \textit{$K$}-matrix is related to the scattering phase
shift, according to
\textit{\eq
\tan\delta_l(q_0)=\dfrac{\mu p}{2\pi}\,K_l(q_0,q_0;z)\, .
\en }
In conclusion, several remarks are in order:
\begin{itemize}
\item[i)]
The decomposition of the two-particle Green's function in a finite volume, given
in eq.~(\ref{eq:split-2}), is valid, if both sides of the same equation are
integrated with the same regular test function. In general, it does not hold,
if integrated with singular functions. In the Born series of the LS equation,
the two-particle Green's function is always integrated with the potential, whose
singularities are assumed to lie far away from the integration contour.

\item[ii)]
If \textit{$\nu^2<0$}, then \textit{$Z_{lm}(1;\nu^2)=-\delta_{l0}\delta_{m_0}\pi^{3/2}
\sqrt{-\nu^2-i0}$}, up to the exponentially suppressed contributions.
This implies that the sum of all the terms, proportional to \textit{$\Phi_0(q_0)$} for 
\textit{$\nu^2<0$} in eq.~(\ref{eq:S2}), is in fact exponentially suppressed.
Hence, the quantity \textit{$S_2$} does not depend on the values
of the function \textit{$\Phi({\bf p})$} outside the original range \textit{$|{\bf p}|>0$}, as it
should.

\item[iii)]
The reason why we still retain these terms and do not replace the quantity
\textit{$\Delta({\bf p}^2,q_0^2)$} by  \textit{$\delta({\bf p}^2-q_0^2)$} everywhere, is the fact
that the principal-value integral, which was defined above, is the regular
function of \textit{$q_0^2$} (no unitary cusp present). 
This property plays no role in the
two-particle scattering (even in the multi-channel case) but, as we shall see below,
 becomes critical
in case of three-particles. Note also that in ref.~\cite{scalar}, 
which deals with the multi-channel scattering for the coupled
$\pi\pi$-$K\bar K$ channels, we have effectively used the same splitting as in 
eq.~(\ref{eq:split-2}) but with \textit{$f(x)=1$}.

\end{itemize}

\section{Three particles in a finite volume}
\label{sec:threebody}

\subsection{Splitting of the three-particle Green's function}
\label{sec:splitting-3}

In analogy with eq.~(\ref{eq:twobody-def-2}) we define the finite-volume 
three-body Green's function in a channel \textit{$\alpha=1,2,3$} as (in order
to ease the notations, we suppress the channel index \textit{$\alpha$} in all momenta)
\textit{\eq\label{eq:threebody-def}
G_{0\alpha}^L({\bf k},{\bf l};z)&=&\dfrac{1}{L^6}\,
\sum_{{\bf p} {\bf q}}
\dfrac{(2\pi)^3\delta^3({\bf p}-{\bf k})(2\pi)^3\delta^3({\bf q}-{\bf l})}
{M+\dfrac{{\bf p}^2}{2M_\alpha}+\dfrac{{\bf q}^2}{2\mu_\alpha}-z}\, ,
\nonumber\\[2mm]
{\bf q}&=&\tilde{\bf p}+\dfrac{m_\beta}{m_\beta+m_\gamma}\,{\bf p}\, ,\quad\quad
({\bf p},\tilde{\bf p})=\dfrac{2\pi}{L}\,({\bf n},\tilde{\bf n})\, ,\quad
{\bf n},\tilde{\bf n}\in\mathbb{Z}^3\, ,
\en }
so that, in the above equation, the summation is carried out over the integers
\textit{${\bf n},\tilde{\bf n}$}.
In order to find the desired form of the splitting, 
we separate the 2-particle Green's function in the above expression
\textit{\eq\label{eq:q02a}
G_{0\alpha}^L({\bf k},{\bf l};z)&=&\dfrac{1}{L^3}\,
\sum_{{\bf p}}(2\pi)^3\delta^3({\bf p}-{\bf k})
\dfrac{2\mu_\alpha}{L^3}\sum_{\bf q}\dfrac{(2\pi)^3\delta^3({\bf q}-{\bf l})}
{{\bf q}^2-q_{0\alpha}^2}\, ,
\nonumber\\[2mm]
q_{0\alpha}^2&=&2\mu_\alpha\biggl(z-M-\dfrac{{\bf p}^2}{2M_\alpha}\biggr)\, .
\en }
Acting in the similar way as in the two-body case (see section~\ref{sec:twobody}),
we obtain
\textit{\eq\label{eq:split3}
G_{0\alpha}^L({\bf k},{\bf l};z)&=&\dfrac{1}{L^3}\,
\sum_{{\bf p}}(2\pi)^3\delta^3({\bf p}-{\bf k})\,\biggl\{
\dfrac{1}{M+\dfrac{{\bf p}^2}{2M_\alpha}+\dfrac{{\bf l}^2}{2\mu_\alpha}-z-i0}
\nonumber\\[2mm]
&-&\,i\pi\Delta\biggl({\bf l}^2,
2\mu_\alpha\biggl(M+\dfrac{{\bf p}^2}{2M_\alpha}-z\biggr)\biggr)
\nonumber\\[2mm]
&+&\Delta\biggl({\bf l}^2,
2\mu_\alpha\biggl(M+\dfrac{{\bf p}^2}{2M_\alpha}-z\biggr)\biggr)
\sum_{lm}Y^*_{lm}(\hat {\bf l})\dfrac{2}{\nu_\alpha^l}Z_{lm}^{\bf a}(1;\nu_\alpha^2)\biggr\}\, ,
\en }
where \textit{$\nu_\alpha=q_{0\alpha}L/(2\pi)$, ${\bf a}=\bigl[m_\beta/(m_\beta+m_\gamma)\bigr]\,
{\bf p}L/(2\pi)$}, and the L\"uscher zeta-function
in the moving frame is defined as (cf. the pertinent expression in the CM frame, eq.~(\ref{eq:zeta}))
\textit{\eq
Z_{lm}^{\bf a}(1;\nu_\alpha^2)=\lim_{\lambda\to\infty}\biggl\{\sum_{{\bf n}\in \mathbb{Z}^3}
\theta(\lambda^2-({\bf n}+{\bf a})^2)\,
\dfrac{{\cal Y}_{lm}({\bf n}+{\bf a})}{({\bf n}+{\bf a})^2-\nu_\alpha^2}-\delta_{l0}\delta_{m0}
\sqrt{4\pi}\lambda\biggr\}\, .
\en }
Since the principal-value integral  over the variable \textit{${\bf l}$}, containing
 a regular function of the arguments
\textit{${\bf p},{\bf l}$},
is a regular function of the remaining variable
\textit{${\bf p}$}, 
one may use the regular summation theorem in this
variable and rewrite eq.~(\ref{eq:split3}) in the following form (cf. 
eq.~(\ref{eq:split-2}))
\textit{\eq\label{eq:split-3}
G_{0\alpha}^L({\bf k},{\bf l};z)&=&
\dfrac{1}{M+\dfrac{{\bf k}^2}{2M_\alpha}+\dfrac{{\bf l}^2}{2\mu_\alpha}-z-i0}
-\,i\pi\Delta\biggl({\bf l}^2,
2\mu_\alpha\biggl(M+\dfrac{{\bf k}^2}{2M_\alpha}-z\biggr)\biggr)
\nonumber\\[2mm]
&+&\dfrac{1}{L^3}\,
\sum_{{\bf p}}(2\pi)^3\delta^3({\bf p}-{\bf k})\,
\Delta\biggl({\bf l}^2,
2\mu_\alpha\biggl(M+\dfrac{{\bf p}^2}{2M_\alpha}-z\biggr)\biggr)
\sum_{lm}Y^*_{lm}(\hat {\bf l})\dfrac{2}{\nu_\alpha^l}Z_{lm}^{\bf a}(1;\nu_\alpha^2)
\nonumber\\[2mm]
&\doteq&
G_{0\alpha}({\bf k},{\bf l};z)+G_{{\sf U}\alpha}({\bf k},{\bf l};z)+G_{{\sf F}\alpha}({\bf k},{\bf l};z)
\nonumber\\[2mm]
&\doteq&
G_{{\sf K}\alpha}({\bf k},{\bf l};z)
+\dfrac{1}{L^3}\,\sum_{\bf p}(2\pi)^3\delta^3({\bf p}-{\bf k})
\tilde G_{{\sf F}\alpha}({\bf p},{\bf l};z)\, .
\en }
Needless to say that the above splitting is valid, if both sides of 
eq.~(\ref{eq:split-3}) are integrated with a regular function of the 
momentum variables \textit{${\bf k}$} and \textit{${\bf l}$}. Further, since the 
variables \textit{${\bf k}$} and \textit{${\bf p}$} are not restricted from above, the
argument of the distribution \textit{$\Delta$} can become positive and arbitrarily large,
independent of the choice of the variables \textit{${\bf l}^2$} and \textit{$z$}. This means that
one has to necessarily deal with the analytic continuation of the pertinent amplitudes
below threshold. Note however that this sub-threshold contribution is effectively
cut by the regulator \textit{$f$} for the momenta 
\textit{$-q_{0\alpha}^2>\mu^2$}. Assuming the range of
the potentials much smaller than the inverse of the lightest mass in the system,
one does not expect to encounter any singularities in the analytic continuation
for \textit{$-q_{0\alpha}^2<\mu^2$}. The above argumentation serves to justify the introduction
of the regulator \textit{$f$}.

Using the representation of the three-particle Green's function, given in 
eq.~(\ref{eq:split-3}), one may try to repeat the same steps as in 
eq.~(\ref{eq:TKTL}), also for the three-particle \textit{$T$}-matrix. In the three-particle
case, however, a new complication arises, related to the presence of the disconnected
contributions (the diagrams where a spectator particle propagates freely,
when the other two particles interact). These diagrams contain the factor
\textit{$L^3\delta_{{\bf p}_\alpha{\bf q}_\alpha}$ ($(2\pi)^3\delta^3({\bf p}_\alpha-{\bf q}_\alpha)$})
in a finite (infinite) volume. This factor is not a regular function and
is \textit{$L$}-dependent (in a finite volume). Consequently, the method, which was used in 
the two-particle case, can be applied here first after these disconnected 
contributions are removed. Below we shall demonstrate, how this can be achieved.

\subsection{Effective two-body \textit{$K$}-matrix in the presence of the \\three-par\-tic\-le intermediate states}

As was shown in section~\ref{sec:twobody}, in order to obtain the L\"uscher
formula from eq.~(\ref{eq:LS-effL}),
one may substitute here the splitting of the two-particle Green's function, given in 
eq.~(\ref{eq:split-2}). The result is given in eq.~(\ref{eq:DM}). The effective
two-body \textit{$K$}-matrix obeys the equation
\textit{\eq\label{eq:K-effL}
\langle {\bf q}'|{\bf K}^L(z)|{\bf q}\rangle=
\langle {\bf q}'|{\bf w}^L(z)|{\bf q}\rangle
+\mbox{\sf P.V.}\int\dfrac{d^3{\bf q}''}{(2\pi)^3}\,
\dfrac{\langle {\bf q}'|{\bf w}^L(z)|{\bf q}''\rangle\,\langle {\bf q}''|{\bf K}^L(z)|{\bf q}\rangle}
{m_1+m_2+\dfrac{({\bf q}'')^2}{2\mu_3}-z}\, .
\en }
Performing the partial-wave expansion of the above equation, we arrive at
eq.~(\ref{eq:KpartialL}). Note that, in difference to eq.~(\ref{eq:K-elastic}), which refers to the elastic case, the effective \textit{$K$}-matrix in the above equation still depends on \textit{$L$}
above the three-particle threshold. 

Taking into account the expression for the potential \textit{${\bf w}^L(z)$}, given in
eq.~(\ref{eq:wL}), we get
\textit{\eq\label{eq:2L}
\langle {\bf q}'|{\bf K}^L(z)|{\bf q}\rangle&=&
\langle {\bf q}'|{\bf K}_3(z)|{\bf q}\rangle
+\dfrac{1}{L^{12}}\,
\sum_{{\bf p}_\alpha''{\bf q}_\alpha''{\bf p}_\beta'''{\bf q}_\beta'''}
\tilde\Gamma_\alpha({\bf q}';{\bf p}_\alpha''{\bf q}_\alpha'')
\nonumber\\[2mm]
&\times&\langle {\bf p}_\alpha''{\bf q}_\alpha''|\tilde{\bf g}^L_3(z)|{\bf p}_\beta'''{\bf q}_\beta'''\rangle
(\tilde\Gamma_\beta({\bf q};{\bf p}_\beta'''{\bf q}_\beta'''))^*\, ,
\en }
where \textit{$K_3$} is defined in the infinite volume
\textit{\eq\label{eq:K3}
\langle {\bf q}'|{\bf K}_3(z)|{\bf q}\rangle=
(-\bar V_3({\bf q}',{\bf q}))
+\mbox{\sf P.V.}\int\dfrac{d^3{\bf q}''}{(2\pi)^3}\,
\dfrac{(-\bar V_3({\bf q}',{\bf q}''))\,\langle {\bf q}''|{\bf K}_3(z)|{\bf q}\rangle}
{m_1+m_2+\dfrac{({\bf q}'')^2}{2\mu_3}-z}\, .
\en }
Further, the quantity \textit{$\tilde\Gamma_\alpha$} is defined through \textit{$\bar\Gamma_\alpha$} in the following manner
\textit{\eq
\tilde\Gamma_\alpha({\bf q}';{\bf p}_\alpha''{\bf q}_\alpha'')
=\bar\Gamma_\alpha({\bf q}';{\bf p}_\alpha''{\bf q}_\alpha'')
+\int\dfrac{d^3{\bf q}''}{(2\pi)^3}\,
\dfrac{\langle{\bf q}'|{\bf K}_3(z)|{\bf q}''\rangle
\bar\Gamma_\alpha({\bf q}'';{\bf p}_\alpha''{\bf q}_\alpha'')}
{m_1+m_2+\dfrac{({\bf q}'')^2}{2\mu_3}-z}\, ,
\en }
and the Green's function \textit{$\tilde{\bf g}^L_3(z)$} obeys the following equation
\textit{\eq
\langle{\bf p}'_\alpha{\bf q}'_\alpha|\tilde{\bf g}^L_3(z)|{\bf p}_\beta{\bf q}_\beta\rangle
=\langle{\bf p}'_\alpha{\bf q}'_\alpha|{\bf g}^L_3(z)|{\bf p}_\beta{\bf q}_\beta\rangle
+\dfrac{1}{L^{12}}\,\sum_{{\bf p}''_\gamma{\bf q}''_\gamma{\bf p}'''_\sigma{\bf q}'''_\sigma}
\langle{\bf p}'_\alpha{\bf q}'_\alpha|{\bf g}^L_3(z)|{\bf p}''_\gamma{\bf q}''_\gamma\rangle
\nonumber\\[2mm]
\times
\langle{\bf p}''_\gamma{\bf q}''_\gamma|(-{\bf V}_4(z))|{\bf p}'''_\sigma{\bf q}'''_\sigma\rangle
\langle{\bf p}'''_\sigma{\bf q}'''_\sigma|\tilde{\bf g}^L_3(z)|{\bf p}_\beta{\bf q}_\beta\rangle\, .
\en }
Here, \textit{${\bf g}^L_3(z)=\sum_{\alpha,\beta=1}^3{\bf g}^L_{3\alpha\beta}(z)$} and \textit{$\gamma,\sigma$} 
can be chosen arbitrarily.
Further, \textit{${\bf V}_4(z)$} denotes the effective three-particle potential, which has emerged
after ``projecting out'' the two-particle intermediate state. This potential
is given by the following expression
\textit{\eq
\langle{\bf p}'_\alpha{\bf q}'_\alpha|(-{\bf V}_4(z))|{\bf p}_\beta{\bf q}_\beta\rangle
&=&\mbox{\sf P.V.}\int\dfrac{d^3{\bf q}'}{(2\pi)^3}\,\dfrac{d^3{\bf q}}{(2\pi)^3}\,
(\bar\Gamma_\alpha({\bf q}';{\bf p}_\alpha'{\bf q}_\alpha'))^*
\biggl\{\dfrac{(2\pi)^3\delta^3({\bf q}'-{\bf q})}{m_1+m_2+\dfrac{{\bf q}^2}{2\mu_3}-z}
\nonumber\\[2mm]
\hspace*{-1.cm}&&\hspace*{-3.cm}+\,\,\dfrac{\langle {\bf q}'|{\bf K}_3(z)|{\bf q}\rangle}{\biggl(m_1+m_2+\dfrac{({\bf q}')^2}{2\mu_3}-z\biggr)
\biggl(m_1+m_2+\dfrac{{\bf q}^2}{2\mu_3}-z\biggr)}
\biggr\}
\bar\Gamma_\beta({\bf q};{\bf p}_\beta{\bf q}_\beta)\, .
\en }
We would like to stress that the effective potential \textit{${\bf V}_4(z)$} is defined
in the infinite volume and is a smooth function of its arguments.

To summarize, after ``projecting out'' the two-particle intermediate state, an
effective three-particle force \textit{${\bf V}_4(z)$} appears in Faddeev equations, even
though we initially assumed that no such force is present. The effective two-body \textit{$K$}-matrix
in a finite volume is given explicitly by eq.~(\ref{eq:2L}). In order to evaluate
the finite-volume effects in the three-body Green's 
function, which enters this expression,
one should consider Faddeev equations in a finite volume.

\subsection{The three-particle counterpart of the L\"uscher formula} 
\label{sec:faddeev}

In order to derive the three-particle analog of the L\"uscher formula,
we have to deal with the three-body equations in the presence 
of the three-particle force. To this end, one may use,
\textit{e.g.}, the formalism described in the papers~\cite{Kowalski-3}
-- with the potentials replaced by the 
\textit{$K$}-matrix elements and the free Green's function replaced by \textit{${\bf G}_{\sf F}$}.
However, as discussed above, there exists a problem related to the presence
of the disconnected parts. Namely, in the presence of the 
Kronecker-\textit{$\delta$}, 
contained in the disconnected parts, the use of the splitting 
procedure for the three-particle propagator according to eq.~(\ref{eq:split-3})
can not be justified mathematically.
In order to circumvent this problem, we act by using a trial and error method.
Namely, we first apply the splitting in the three-body LS equations,
as if there were no disconnected parts, and further use the Faddeev trick. 
In the resulting equations,
the disconnected terms, containing the \textit{$\delta$}-functions, emerge in a finite
volume. At the next stage, we discard these singular terms by hand, thus 
making a {\em conjecture} about the correct form of the equations. 
At the final step, we check this conjecture
explicitly, by considering the 
multiple-scattering series that emerge from the resulting equations,
 and showing that
this series coincides with the original multiple-scattering 
series in a finite volume. 

\begin{figure}[t]
\begin{center}
\includegraphics[width=8.cm]{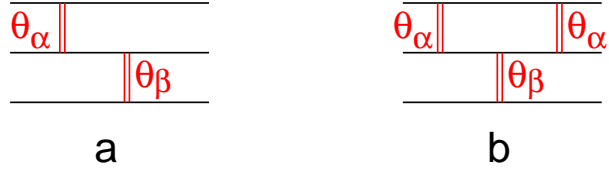}
\end{center}
\caption{a) One iteration of the disconnected diagrams produces a connected
diagram with no loops; b) Loops emerge first after two iterations.}
\label{fig:iterations}
\end{figure}

Symbolically, the result for the effective two-body \textit{$K$}-matrix can be written 
as follows
\textit{\eq
{\bf K}^L={\bf K}_{2\to 2}+{\bf K}_{2\to 3}({\bf G}_{\sf F}
+{\bf G}_{\sf F}{\bf R}_{\sf F}{\bf G}_{\sf F}){\bf K}_{3\to 2}\, ,
\en }
where \textit{${\bf K}_{i\to j}$} denote the pertinent \textit{$K$}-matrix elements in 
the infinite volume\footnote{Above threshold, the three-body \textit{$K$}-matrix
 coincides with the one defined, \textit{e.g.}, in refs.~\cite{Manning,Kowalski-K}.}, 
\textit{${\bf G}_{\sf F}$} stands for the finite-volume part of the 
three-particle Green's function (see eq.~(\ref{eq:split-3})), and the quantity 
\textit{${\bf R}_{\sf F}$} is given by
\textit{\eq\label{eq:R}
{\bf R}_{\sf F}&=&\sum_{\mu,\nu=1}^4{\bf R}_{\mu\nu}+\sum_{\alpha=1}^3\btheta_\alpha\, ,
\nonumber\\[2mm]
{\bf R}_{4\beta}&=&\btheta_4{\bf G}_{\sf F}\biggl(\btheta_\beta
+\sum_{\gamma=1}^3{\bf R}_{\gamma\beta}\biggr)\, ,
\nonumber\\[2mm]
{\bf R}_{\alpha\beta}&=&\btheta_\alpha{\bf G}_{\sf F}
\biggl(\sum_{\gamma=1}^3(1-\delta_{\alpha\gamma}){\bf R}_{\gamma\beta}+{\bf R}_{4\beta}
\biggr)\, ,
\nonumber\\[2mm]
{\bf R}_{\alpha4}&=&\btheta_\alpha{\bf G}_{\sf F}\sum_{\gamma=1}^3(1-\delta_{\alpha\gamma})
{\bf R}_{\gamma4}+\btheta_\alpha{\bf G}_{\sf F}{\bf R}_{44}\, ,
\nonumber\\[2mm]
{\bf R}_{44}&=&\btheta_4+\btheta_4{\bf G}_{\sf F}\sum_{\gamma=1}^3{\bf R}_{\gamma4}
\, .
\en }
Here, \textit{$\mu,\nu=1,\cdots, 4$}, whereas \textit{$\alpha,\beta,\gamma=1,\cdots, 3$} and
\textit{\eq
\btheta_\mu={\bf K}_\mu+{\bf K}_\mu{\bf G}_{\sf F}\btheta_\mu\, ,\quad\quad
{\bf K}_\mu=(-{\bf V}_\mu)+(-{\bf V}_\mu)({\bf G}_{\sf 0}+{\bf G}_{\sf U}){\bf K}_\mu\, ,\quad\quad
{\bf K}_{3\to 3}=\sum_{\mu=1}^4{\bf K}_\mu\, .
\en }
Note that in eq.~(\ref{eq:R}) (third line), we have omitted the terms 
of the type \textit{$\btheta_\alpha{\bf G}_{\sf F}\btheta_\beta$} with \textit{$\alpha\neq\beta$}.
Physically, such terms correspond to the finite-volume corrections in the
 disconnected diagrams and emerge, if one faithfully
applies the splitting procedure even to the disconnected piece.
This omission will be justified below by showing that eq.~(\ref{eq:R})
produces -- diagram by diagram -- 
the correct splitting of the infinite- and finite-volume parts 
in the multiple-scattering series\footnote{The physical meaning of this
 prescription is very transparent. The finite-volume corrections emerge only
in the loop diagrams. However, due to the presence of the Kronecker-delta 
in the disconnected diagrams, a first iteration of the disconnected diagrams 
in the Faddeev equations gives a connected diagram without a loop. Its explicit expression is identical 
in a finite and the infinite volumes. The loops (and, consequently, 
the finite-volume corrections) emerge first in the second iteration, see fig.~\ref{fig:iterations}}.
Here we also would like to mention that this system of equations is
formally identical to the equations that relate the \textit{$T$}-matrix and \textit{$K$}-matrix
elements, with a replacement \textit{${\bf R}\to {\bf K}$, ${\bf K}\to{\bf T}$}
and \textit{${\bf G}_{\sf F}\to{\bf G}_{\sf U}$}.

The above equations can be written down in the explicit form 
\textit{\eq\label{eq:KL}
\langle {\bf q}'|{\bf K}^L|{\bf q}\rangle=
\langle {\bf q}'|{\bf K}_{2\to 2}|{\bf q}\rangle
+\langle {\bf q}'|{\bf K}_{2\to 3}{\bf G}_{\sf F}{\bf K}_{3\to 2}|{\bf q}\rangle
+\langle {\bf q}'|{\bf K}_{2\to 3}{\bf G}_{\sf F}{\bf R}_{\sf F}{\bf G}_{\sf F}{\bf K}_{3\to 2}|{\bf q}\rangle\, ,
\en }
where\footnote{Note the superscript \textit{$(\alpha)$} in the \textit{$K$}-matrix elements. 
It emerges because \textit{${\bf G}_{{\sf U}\alpha}$} explicitly depends on the channel
\textit{$\alpha$} through the regulator \textit{$f(q_{0\alpha}^2/\mu^2)$}, see
eq.~(\ref{eq:split-3}).}
\textit{\eq
\langle {\bf q}'|{\bf K}_{2\to 3}{\bf G}_{\sf F}{\bf K}_{3\to 2}|{\bf q}\rangle
=\dfrac{1}{L^3}\sum_{{\bf p}_\alpha}
\int\dfrac{d^3{\bf q}_\alpha}{(2\pi)^3}\,
\langle {\bf q}'|{\bf K}^{(\alpha)}_{2\to 3}|{\bf p}_\alpha{\bf q}_\alpha\rangle
\tilde G_{{\sf F}\alpha}({\bf p}_\alpha,{\bf q}_\alpha;z)
\langle {\bf p}_\alpha{\bf q}_\alpha|{\bf K}^{(\alpha)}_{3\to 2}|{\bf q}\rangle\, ,
\en }
The last term in eq.~(\ref{eq:KL}) is given by
\textit{\eq
\langle {\bf q}'|{\bf K}_{2\to 3}{\bf G}_{\sf F}{\bf R}_{\sf F}
{\bf G}_{\sf F}{\bf K}_{3\to 2}|{\bf q}\rangle
&=&\langle {\bf q}'|{\bf K}_{2\to 3}{\bf G}_{\sf F}\biggl(\sum_{\alpha=1}^3
\btheta_\alpha+\sum_{\mu,\nu=1}^4{\bf R}_{\mu\nu}\biggr)
{\bf G}_{\sf F}{\bf K}_{3\to 2}|{\bf q}\rangle\, ,
\en }
where
\textit{\eq
\langle {\bf q}'|{\bf K}_{2\to 3}{\bf G}_{\sf F}\btheta_\alpha
{\bf G}_{\sf F}{\bf K}_{3\to 2}|{\bf q}\rangle
&=&\dfrac{1}{L^3}\sum_{{\bf p}_\alpha}\int\dfrac{d^3{\bf q}'_\alpha}{(2\pi)^3}
\dfrac{d^3{\bf q}_\alpha}{(2\pi)^3}\,
\langle {\bf q}'|{\bf K}^{(\alpha)}_{2\to 3}|{\bf p}_\alpha{\bf q}'_\alpha\rangle
\tilde G_{{\sf F}\alpha}({\bf p}_\alpha,{\bf q}'_\alpha;z)
\nonumber\\[2mm]
&\times&
\langle {\bf q}'_\alpha|\btheta_\alpha(z;{\bf p}_\alpha)|{\bf q}_\alpha\rangle
\tilde G_{{\sf F}\alpha}({\bf p}_\alpha,{\bf q}_\alpha;z)
\langle {\bf p}_\alpha{\bf q}_\alpha|{\bf K}^{(\alpha)}_{3\to 2}|{\bf q}\rangle
\en }
and
\textit{\eq
\langle {\bf q}'|{\bf K}_{2\to 3}{\bf G}_{\sf F}{\bf R}_{\mu\nu}
{\bf G}_{\sf F}{\bf K}_{3\to 2}|{\bf q}\rangle
&=&\dfrac{1}{L^6}\sum_{{\bf p}'_\alpha{\bf p}_\beta}
\int\dfrac{d^3{\bf q}'_\alpha}{(2\pi)^3}
\dfrac{d^3{\bf q}_\beta}{(2\pi)^3}\,
\langle {\bf q}'|{\bf K}^{(\alpha)}_{2\to 3}|{\bf p}'_\alpha{\bf q}'_\alpha\rangle
\tilde G_{{\sf F}\alpha}({\bf p}'_\alpha,{\bf q}'_\alpha;z)
\nonumber\\[2mm]
&\times&\langle {\bf p}'_\alpha{\bf q}'_\alpha|{\bf R}_{\mu\nu}|{\bf p}_\beta{\bf q}_\beta\rangle
\tilde G_{{\sf F}\beta}({\bf p}_\beta,{\bf q}_\beta;z)
\langle {\bf p}_\beta{\bf q}_\beta|{\bf K}^{(\beta)}_{3\to 2}|{\bf q}\rangle\, .
\en }
Note that if \textit{$\mu=4$} and/or \textit{$\nu=4$} above, the pertinent 
indexes \textit{$\alpha,\beta$} can be chosen arbitrarily.

Next, \textit{$\btheta_\alpha$} are determined through
\textit{\eq\label{eq:subsystem}
&&\langle {\bf q}'_\alpha|\btheta_\alpha(z;{\bf p}_\alpha)|{\bf q}_\alpha\rangle
=\langle {\bf q}'_\alpha|{\bf K}_\alpha\biggl(z-m_\alpha
-\dfrac{{\bf p}_\alpha^2}{2M_\alpha}\biggr)|{\bf q}_\alpha\rangle
\nonumber\\[2mm]
&+&\int\dfrac{d^3{\bf l}_\alpha}{(2\pi)^3}\,\langle {\bf q}'_\alpha|{\bf K}_\alpha\biggl(z-m_\alpha-\dfrac{{\bf p}_\alpha^2}{2M_\alpha}\biggr)|{\bf l}_\alpha\rangle
\tilde G_{{\sf F}\alpha}({\bf p}_\alpha,{\bf l}_\alpha;z)
\langle {\bf l}_\alpha|\btheta_\alpha(z;{\bf p}_\alpha)|{\bf q}_\alpha\rangle\, .
\en }
In this expression, \textit{${\bf K}_\alpha$} stands for the {\em two-particle} \textit{$K$}-matrix.
Performing partial-wave expansion
in this equation and integrating over the absolute value \textit{$|{\bf l}_\alpha|$}, we
arrive at a set of {\em algebraic} equations that relate \textit{${\bf K}_\alpha$}
and \textit{$\btheta_\alpha$}. These equations mix all partial waves. Hence, 
in order to solve them, a partial-wave truncation is necessary. It is easy
to recognize that eq.~(\ref{eq:subsystem}) is nothing but the L\"uscher formula
for a two-particle sub-system in the moving frame (the third particle plays
the role of a spectator). The result was of course expected from the beginning.

The equation for \textit{$\btheta_4$} is given by
\textit{\eq\label{eq:subsystem-4}
\langle {\bf p}'_\alpha{\bf q}'_\alpha|\btheta_4|{\bf p}_\beta{\bf q}_\beta\rangle
&=&
\langle {\bf p}'_\alpha{\bf q}'_\alpha|{\bf K}_4|{\bf p}_\beta{\bf q}_\beta\rangle
\nonumber\\[2mm]
&+&\dfrac{1}{L^3}\sum_{{\bf k}_\alpha}
\int\dfrac{d^3{\bf l}_\alpha}{(2\pi)^3}\,
\langle {\bf p}'_\alpha{\bf q}'_\alpha|{\bf K}_4|{\bf k}_\alpha{\bf l}_\alpha\rangle
\tilde G_{{\sf F}\alpha}({\bf k}_\alpha,{\bf l}_\alpha;z)
\langle {\bf k}_\alpha{\bf l}_\alpha|\btheta_4|{\bf p}_\beta{\bf q}_\beta\rangle\, .
\en }
It is easy to see that, after performing the integration over the absolute value
of the relative momentum \textit{${\bf l}_\alpha$}, 
eq.~(\ref{eq:subsystem-4}) transforms into the 
matrix equation that can be solved straightforwardly by using numerical 
methods. The matrix indexes
here include the discrete values of the spectator 
momenta \textit{${\bf p}_\alpha,{\bf q}_\beta,{\bf k}_\alpha$}. 
In order to solve this equation,
a method analogous to the one described in ref.~\cite{Doring:2011ip} can be used.

Finally, we write down the equations~(\ref{eq:R}) in the explicit form 
\textit{\eq
\langle {\bf p}'_\alpha{\bf q}'_\alpha|{\bf R}_{4\beta}|{\bf p}_\beta{\bf q}_\beta\rangle
=\int\dfrac{d^3{\bf q}_\beta''}{(2\pi)^3}\,
\langle {\bf p}'_\alpha{\bf q}'_\alpha|\btheta_4|{\bf p}_\beta{\bf q}''_\beta\rangle
\tilde G_{{\sf F}\beta}({\bf p}_\beta,{\bf q}''_\beta;z)
\langle{\bf q}''_\beta|\btheta_\beta(z;{\bf p}_\beta)|{\bf q}_\beta\rangle
\nonumber\\[2mm]
+\sum_{\gamma=1}^3\dfrac{1}{L^3}\sum_{{\bf p}''_\gamma}
\int\dfrac{d^3{\bf q}_\gamma''}{(2\pi)^3}\,
\langle {\bf p}'_\alpha{\bf q}'_\alpha|\btheta_4|{\bf p}''_\gamma{\bf q}''_\gamma\rangle
\tilde G_{{\sf F}\gamma}({\bf p}''_\gamma,{\bf q}''_\gamma;z)
\langle{\bf p}''_\gamma{\bf q}''_\gamma|{\bf R}_{\gamma\beta}|{\bf p}_\beta{\bf q}_\beta\rangle\, ,
\en }
\textit{\eq
\langle {\bf p}'_\alpha{\bf q}'_\alpha|{\bf R}_{\alpha\beta}|{\bf p}_\beta{\bf q}_\beta\rangle
&=&\sum_{\gamma=1}^3(1-\delta_{\alpha\gamma})\int\dfrac{d^3{\bf q}''_\alpha}{(2\pi)^3}\,
\langle{\bf q}'_\alpha|\btheta({\bf p}'_\alpha;z)|{\bf q}''_\alpha\rangle
\tilde G_{{\sf F}\alpha}({\bf p}'_\alpha,{\bf q}''_\alpha;z)
\langle \bar{\bf p}''_\gamma\bar{\bf q}''_\gamma|
{\bf R}_{\gamma\beta}|{\bf p}_\beta{\bf q}_\beta\rangle
 \nonumber\\[2mm]
&+&\int\dfrac{d^3{\bf q}''_\alpha}{(2\pi)^3}\,\langle{\bf q}'_\alpha|\btheta({\bf p}'_\alpha;z)|{\bf q}''_\alpha\rangle
\tilde G_{{\sf F}\alpha}({\bf p}'_\alpha,{\bf q}''_\alpha;z)
\langle {\bf p}'_\alpha{\bf q}''_\alpha|
{\bf R}_{4\beta}|{\bf p}_\beta{\bf q}_\beta\rangle\, ,
\en }
\textit{\eq
\langle {\bf p}'_\alpha{\bf q}'_\alpha|{\bf R}_{\alpha4}|{\bf p}_\beta{\bf q}_\beta\rangle
&=&\sum_{\gamma=1}^3(1-\delta_{\alpha\gamma})\int\dfrac{d^3{\bf q}''_\alpha}{(2\pi)^3}\,
\langle{\bf q}'_\alpha|\btheta({\bf p}'_\alpha;z)|{\bf q}''_\alpha\rangle
\tilde G_{{\sf F}\alpha}({\bf p}'_\alpha,{\bf q}''_\alpha;z)
\langle \bar{\bf p}''_\gamma\bar{\bf q}''_\gamma|
{\bf R}_{\gamma4}|{\bf p}_\beta{\bf q}_\beta\rangle
 \nonumber\\[2mm]
&+&\int\dfrac{d^3{\bf q}''_\alpha}{(2\pi)^3}\,\langle{\bf q}'_\alpha|\btheta({\bf p}'_\alpha;z)|{\bf q}''_\alpha\rangle
\tilde G_{{\sf F}\alpha}({\bf p}'_\alpha,{\bf q}''_\alpha;z)
\langle {\bf p}'_\alpha{\bf q}''_\alpha|
{\bf R}_{44}|{\bf p}_\beta{\bf q}_\beta\rangle\, ,
\en }
\textit{\eq
&&\langle {\bf p}'_\alpha{\bf q}'_\alpha|{\bf R}_{44}|{\bf p}_\beta{\bf q}_\beta\rangle
=\langle {\bf p}'_\alpha{\bf q}'_\alpha|\btheta_{44}|{\bf p}_\beta{\bf q}_\beta\rangle
\nonumber\\[2mm]
 &+&\sum_{\gamma=1}^3\dfrac{1}{L^3}\sum_{{\bf p}''_\gamma}
\int\dfrac{d^3{\bf q}''_\gamma}{(2\pi)^3}
\langle {\bf p}'_\alpha{\bf q}'_\alpha|\btheta_{44}|{\bf p}''_\gamma{\bf q}''_\gamma\rangle
\tilde G_{{\sf F}\gamma}({\bf p}''_\gamma,{\bf q}''_\gamma;z)
\langle {\bf p}''_\gamma{\bf q}''_\gamma|{\bf R}_{\gamma4}|{\bf p}_\beta{\bf q}_\beta\rangle\, ,
\en }
where
\textit{\eq
\bar{\bf p}''_\gamma=-\dfrac{m_\gamma}{m_\beta+m_\gamma}\,{\bf p}'_\alpha-{\bf q}''_\alpha
\, ,\quad\quad
\bar{\bf q}''_\gamma=\dfrac{m_\beta(m_\alpha+m_\beta+m_\gamma)}{(m_\alpha+m_\beta)(m_\beta+m_\gamma)}\,
{\bf p}'_\alpha-\dfrac{m_\alpha}{m_\alpha+m_\beta}\,{\bf q}''_\alpha\, .
\en }

\subsection{The term without a potential insertion}
\label{sec:no}

Below we shall start  examining the multiple scattering series 
which are obtained from eq.~(\ref{eq:2L})
diagram by diagram, and shall
demonstrate -- for a few illuminating cases -- that our equations
 indeed produce the desired splitting of the infinite- and finite-volume
contributions.

We start from the trivial case of a diagram without rescattering, shown
in fig.~\ref{fig:g3Lz}a. This diagram should be folded with the vertex functions
\textit{$\tilde\Gamma,\tilde\Gamma^*$}. As a result, we arrive at the following
expression
\textit{\eq
I_0=\dfrac{1}{L^6}\, \sum_{{\bf p}_3{\bf q}_3}
\dfrac{\tilde\Gamma_3({\bf q}';{\bf p}_3{\bf q}_3)
(\tilde\Gamma_3({\bf q};{\bf p}_3{\bf q}_3))^*}
{M+\dfrac{{\bf p}_3^2}{2M_3}+\dfrac{{\bf q}_3^2}{2\mu_3}-z}\, .
\en }
This expression is of the type already considered in 
section~\ref{sec:splitting-3}, since the product of two vertices is a regular
function of the momenta. Consequently, up to exponentially suppressed terms,
\textit{\eq
I_0&=&{\sf P.V.}\int\dfrac{d^3{\bf p}_3}{(2\pi)^3}\dfrac{d^3{\bf q}_3}{(2\pi)^3}\,
\dfrac{\tilde\Gamma_3({\bf q}';{\bf p}_3{\bf q}_3)
(\tilde\Gamma_3({\bf q};{\bf p}_3{\bf q}_3))^*}
{M+\dfrac{{\bf p}_3^2}{2M_3}+\dfrac{{\bf q}_3^2}{2\mu_3}-z}
\nonumber\\[2mm]
&+&\dfrac{1}{L^3}\,\sum_{{\bf p}_3}\int\dfrac{d^3{\bf q}_3}{(2\pi)^3}\,
\tilde\Gamma_3({\bf q}';{\bf p}_3{\bf q}_3)
\tilde G_{{\sf F}3}({\bf p}_3,{\bf q}_3;z)
(\tilde\Gamma_3({\bf q};{\bf p}_3{\bf q}_3))^*\, .
\en }

\subsection{Disconnected contributions}
\label{sec:spectator}

The corresponding diagram is shown in fig.~\ref{fig:g3Lz}b. Expanding 
\textit{$\btau_\alpha$} in Born series, one obtains the diagrams with one, two, \textit{$\ldots$}
insertions of the potential. The explicit expression of a diagram
with one insertion is given by
\textit{\eq
I_1^d=\sum_{\alpha=1}^3\dfrac{1}{L^9}\, \sum_{{\bf p}_\alpha{\bf q}'_\alpha{\bf q}_\alpha}
\dfrac{\tilde\Gamma_\alpha({\bf q}';{\bf p}_\alpha{\bf q}'_\alpha)
(-\bar V_\alpha({\bf q}'_\alpha,{\bf q}_\alpha))
(\tilde\Gamma_\alpha({\bf q};{\bf p}_\alpha{\bf q}_\alpha))^*}
{\biggl(M+\dfrac{{\bf p}_\alpha^2}{2M_\alpha}+\dfrac{({\bf q}'_\alpha)^2}{2\mu_\alpha}-z\biggr)\biggl(M+\dfrac{{\bf p}_\alpha^2}{2M_\alpha}+\dfrac{{\bf q}_\alpha^2}{2\mu\alpha}-z\biggr)}\, .
\en }
Carrying out the splitting of the infinite- and finite-volume contributions
explicitly, it is easy to see that, up to the exponentially suppressed
terms, the above equation can be rewritten as
\textit{\eq
I_1^d&=&\sum_{\alpha=1}^3\int\dfrac{d^3{\bf p}_\alpha}{(2\pi)^3}
\dfrac{d^3{\bf q}'_\alpha}{(2\pi)^3}\dfrac{d^3{\bf q}_\alpha}{(2\pi)^3}\,
\tilde\Gamma_\alpha({\bf q}';{\bf p}_\alpha{\bf q}'_\alpha)G_{{\sf K}\alpha}({\bf p}_\alpha,{\bf q}'_\alpha;z)
(-\bar V_\alpha({\bf q}'_\alpha,{\bf q}_\alpha))
\nonumber\\[2mm]
&&\hspace*{3.cm}\times\,\,G_{{\sf K}\alpha}({\bf p}_\alpha,{\bf q}_\alpha;z)
(\tilde\Gamma_\alpha({\bf q};{\bf p}_\alpha{\bf q}_\alpha))^*
\nonumber\\[2mm]
&+&\sum_{\alpha=1}^3\dfrac{1}{L^3}\,\sum_{{\bf p}_\alpha}
\int\dfrac{d^3{\bf q}'_\alpha}{(2\pi)^3}\dfrac{d^3{\bf q}_\alpha}{(2\pi)^3}\,
\tilde\Gamma_\alpha({\bf q}';{\bf p}_\alpha{\bf q}_\alpha)
U^d_{1\alpha}({\bf q}'_\alpha,{\bf q}_\alpha;{\bf p}_\alpha,z)
(\tilde\Gamma_\alpha({\bf q};{\bf p}_\alpha{\bf q}_\alpha))^*\, ,
\en }
where
\textit{\eq
U^d_{1\alpha}({\bf q}'_\alpha,{\bf q}_\alpha;{\bf p}_\alpha,z)
&=&G_{{\sf K}\alpha}({\bf p}_\alpha,{\bf q}'_\alpha;z)
(-\bar V_\alpha({\bf q}'_\alpha,{\bf q}_\alpha))
\tilde G_{{\sf F}\alpha}({\bf p}_\alpha,{\bf q}_\alpha;z)
\nonumber\\[2mm]
&+&\tilde G_{{\sf F}\alpha}({\bf p}_\alpha,{\bf q}'_\alpha;z)
(-\bar V_\alpha({\bf q}'_\alpha,{\bf q}_\alpha))
G_{{\sf K}\alpha}({\bf p}_\alpha,{\bf q}_\alpha;z)
\nonumber\\[2mm]
&+&\tilde G_{{\sf F}\alpha}({\bf p}_\alpha,{\bf q}'_\alpha;z)
(-\bar V_\alpha({\bf q}'_\alpha,{\bf q}_\alpha))
\tilde G_{{\sf F}\alpha}({\bf p}_\alpha,{\bf q}_\alpha;z)\, .
\en }
The generalization for the diagrams with many potential insertions between the
same two particles is straightforward. Further, recalling the definition
of \textit{${\bf K}_{2\to 3}$} and \textit{${\bf K}_{3\to 2}$}, it is easy to ensure that
the diagrams considered in sections~\ref{sec:no} and \ref{sec:spectator},
can be unambiguously identified with the pertinent diagrams 
emerging in the perturbative expansion of
the quantity \textit{${\bf K}_{2\to 2}+{\bf K}_{2\to 3}({\bf G}_{\sf F}
+{\bf G}_{\sf F}\sum_{\alpha=1}^3
\btheta_\alpha{\bf G}_{\sf F}){\bf K}_{3\to 2}$}.

\subsection{Connected contributions}
\label{sec:connected}

The connected contributions emerge as a result of the expansion of the 
diagrams shown in fig.~\ref{fig:g3Lz}c in powers of the potentials
\textit{$\bar V_\alpha$}. The lowest-order term has \textit{$O(\bar V_\alpha^2)$} and is given by
\textit{\eq
I_2^c&=&\sum_{\alpha\neq\beta}\dfrac{1}{L^{12}}
\sum_{{\bf p}'_\alpha{\bf q}'_\alpha{\bf p}_\beta{\bf q}_\beta}
\dfrac{\tilde\Gamma_\alpha({\bf q}';{\bf p}'_\alpha{\bf q}'_\alpha)
(-\bar V_\alpha({\bf q}'_\alpha,{\bf p}_\beta+\dfrac{m_\beta}{M-m_\alpha}\,
{\bf p}'_\alpha))}
{\biggl(M+\dfrac{({\bf p}'_\alpha)^2}{2M_\alpha}
+\dfrac{({\bf q}'_\alpha)^2}{2\mu_\alpha}-z\biggr)
\biggl(M+\dfrac{({\bf p}'_\alpha)^2}{2M_\alpha}
+\dfrac{({\bf p}_\beta+\dfrac{m_\beta}{M-m_\alpha}\,
{\bf p}'_\alpha)^2}{2\mu_\alpha}-z\biggr)}
\nonumber\\[2mm]
&\times&\dfrac{(-\bar V_\beta(-{\bf p}'_\alpha- \dfrac{m_\alpha}{M-m_\beta}\,
{\bf p}_\beta,{\bf q}_\beta))
(\tilde\Gamma_\beta({\bf q};{\bf p}_\beta{\bf q}_\beta))^*}
{\biggl(M+\dfrac{{\bf p}_\beta^2}{2M_\beta}
+\dfrac{{\bf q}_\beta^2}{2\mu_\beta}-z\biggr)}\, .
\en } 
Up to the exponentially suppressed terms, this expression can be rewritten as
\textit{\eq
I_2^c&=&\sum_{\alpha\neq\beta}
\int\dfrac{d^3{\bf p}'_\alpha}{(2\pi)^3}\dfrac{d^3{\bf q}'_\alpha}{(2\pi)^3}
\dfrac{d^3{\bf p}_\beta}{(2\pi)^3}\dfrac{d^3{\bf q}_\beta}{(2\pi)^3}\,
\tilde\Gamma_\alpha({\bf q}';{\bf p}'_\alpha{\bf q}'_\alpha)
U^{c,0}_{2\alpha\beta}({\bf p}'_\alpha{\bf q}'_\alpha;{\bf p}_\beta{\bf q}_\beta)
(\tilde\Gamma_\beta({\bf q};{\bf p}_\beta{\bf q}_\beta))^*
\nonumber\\[2mm]
&+&\sum_{\alpha\neq\beta}\dfrac{1}{L^3}\sum_{{\bf p}'_\alpha}
\int\dfrac{d^3{\bf q}'_\alpha}{(2\pi)^3}
\dfrac{d^3{\bf p}_\beta}{(2\pi)^3}\dfrac{d^3{\bf q}_\beta}{(2\pi)^3}\,
\tilde\Gamma_\alpha({\bf q}';{\bf p}'_\alpha{\bf q}'_\alpha)
U^{c,1}_{2\alpha\beta}({\bf p}'_\alpha{\bf q}'_\alpha;{\bf p}_\beta{\bf q}_\beta)
(\tilde\Gamma_\beta({\bf q};{\bf p}_\beta{\bf q}_\beta))^*
\nonumber\\[2mm]
&+&\sum_{\alpha\neq\beta}\dfrac{1}{L^3}\sum_{{\bf p}_\beta}
\int\dfrac{d^3{\bf p}'_\alpha}{(2\pi)^3}\dfrac{d^3{\bf q}'_\alpha}{(2\pi)^3}
\dfrac{d^3{\bf q}_\beta}{(2\pi)^3}\,
\tilde\Gamma_\alpha({\bf q}';{\bf p}'_\alpha{\bf q}'_\alpha)
U^{c,2}_{2\alpha\beta}({\bf p}'_\alpha{\bf q}'_\alpha;{\bf p}_\beta{\bf q}_\beta)
(\tilde\Gamma_\beta({\bf q};{\bf p}_\beta{\bf q}_\beta))^*
\nonumber\\[2mm]
&+&\sum_{\alpha\neq\beta}\dfrac{1}{L^6}\sum_{{\bf p}'_\alpha{\bf p}_\beta}
\int\dfrac{d^3{\bf q}'_\alpha}{(2\pi)^3}
\dfrac{d^3{\bf q}_\beta}{(2\pi)^3}\,
\tilde\Gamma_\alpha({\bf q}';{\bf p}'_\alpha{\bf q}'_\alpha)
U^{c,3}_{2\alpha\beta}({\bf p}'_\alpha{\bf q}'_\alpha;{\bf p}_\beta{\bf q}_\beta)
(\tilde\Gamma_\beta({\bf q};{\bf p}_\beta{\bf q}_\beta))^*
\, ,
\en }
where
\textit{\eq\label{eq:V20}
U^{c,0}_{2\alpha\beta}({\bf p}'_\alpha{\bf q}'_\alpha;{\bf p}_\beta{\bf q}_\beta)
&\!\!\!=\!\!\!&G_{{\sf K}\alpha}({\bf p}'_\alpha,{\bf q}'_\alpha;z)
(-\bar V_\alpha({\bf q}'_\alpha,\bar {\bf q}''_\alpha))
G_{{\sf K}\alpha}({\bf p}'_\alpha,\bar {\bf q}''_\alpha;z)
(-\bar V_\beta(\bar {\bf q}'''_\beta,{\bf q}_\beta))
G_{{\sf K}\beta}({\bf p}_\beta,{\bf q}_\beta;z) ,
\nonumber\\
&&
\en }
\textit{\eq
U^{c,1}_{2\alpha\beta}({\bf p}'_\alpha{\bf q}'_\alpha;{\bf p}_\beta{\bf q}_\beta)
&\!\!\!=\!\!\!&
G_{{\sf K}\alpha}({\bf p}'_\alpha,{\bf q}'_\alpha;z)
(-\bar V_\alpha({\bf q}'_\alpha,\bar {\bf q}''_\alpha))
\tilde G_{{\sf F}\alpha}({\bf p}'_\alpha,\bar {\bf q}''_\alpha;z)
(-\bar V_\beta(\bar {\bf q}'''_\beta,{\bf q}_\beta))
G_{{\sf K}\beta}({\bf p}_\beta,{\bf q}_\beta;z)
\nonumber\\[2mm]
&\!\!\!+\!\!\!&\tilde G_{{\sf F}\alpha}({\bf p}'_\alpha,{\bf q}'_\alpha;z)
(-\bar V_\alpha({\bf q}'_\alpha,\bar {\bf q}''_\alpha))
G_{{\sf K}\alpha}({\bf p}'_\alpha,\bar {\bf q}''_\alpha;z)
(-\bar V_\beta(\bar {\bf q}'''_\beta,{\bf q}_\beta))
G_{{\sf K}\beta}({\bf p}_\beta,{\bf q}_\beta;z)
\nonumber\\[2mm]
&\!\!\!+\!\!\!&\tilde G_{{\sf F}\alpha}({\bf p}'_\alpha,{\bf q}'_\alpha;z)
(-\bar V_\alpha({\bf q}'_\alpha,\bar {\bf q}''_\alpha))
\tilde G_{{\sf F}\alpha}({\bf p}'_\alpha,\bar {\bf q}''_\alpha;z)
(-\bar V_\beta(\bar {\bf q}'''_\beta,{\bf q}_\beta))
G_{{\sf K}\beta}({\bf p}_\beta,{\bf q}_\beta;z)\, ,
\nonumber\\
&&
\en }
\textit{\eq
U^{c,2}_{2\alpha\beta}({\bf p}'_\alpha{\bf q}'_\alpha;{\bf p}_\beta{\bf q}_\beta)
&\!\!\!=\!\!\!&
\,G_{{\sf K}\alpha}({\bf p}'_\alpha,{\bf q}'_\alpha;z)
(-\bar V_\alpha({\bf q}'_\alpha,\bar {\bf q}''_\alpha))
G_{{\sf K}\beta}({\bf p}_\beta,\bar {\bf q}'''_\beta;z)
(-\bar V_\beta(\bar {\bf q}'''_\beta,{\bf q}_\beta))
\tilde G_{{\sf F}\beta}({\bf p}_\beta,{\bf q}_\beta;z)
\nonumber\\[2mm]
&\!\!\!+\!\!\!&G_{{\sf K}\alpha}({\bf p}'_\alpha,{\bf q}'_\alpha;z)
(-\bar V_\alpha({\bf q}'_\alpha,\bar {\bf q}''_\alpha))
\tilde G_{{\sf F}\beta}({\bf p}_\beta,\bar {\bf q}'''_\beta;z)
(-\bar V_\beta(\bar {\bf q}'''_\beta,{\bf q}_\beta))
\tilde G_{{\sf F}\beta}({\bf p}_\beta,{\bf q}_\beta;z) ,
\nonumber\\
&&
\en }
\textit{\eq
\label{eq:V23}
U^{c,3}_{2\alpha\beta}({\bf p}'_\alpha{\bf q}'_\alpha;{\bf p}_\beta{\bf q}_\beta)
&\!\!=\!\!&
\tilde G_{{\sf F}\alpha}({\bf p}'_\alpha,{\bf q}'_\alpha;z)
\dfrac{(-\bar V_\alpha({\bf q}'_\alpha,\bar {\bf q}''_\alpha))
(-\bar V_\beta(\bar {\bf q}'''_\beta,{\bf q}_\beta))}
{\biggl(M+\dfrac{{\bf p}_\beta^2}{2M_\beta}
+\dfrac{(\bar {\bf q}'''_\beta)^2}{2\mu_\beta}-z\biggr)}
\tilde G_{{\sf F}\beta}({\bf p}_\beta,{\bf q}_\beta;z)\, ,
\nonumber\\
&&
\en }
and
\textit{\eq
\bar {\bf q}''_\alpha={\bf p}_\beta+\dfrac{m_\beta}{M-m_\alpha}\,{\bf p}'_\alpha\, ,\quad\quad
\bar {\bf q}'''_\beta=-{\bf p}'_\alpha- \dfrac{m_\alpha}{M-m_\beta}\,{\bf p}_\beta\, .
\en }
It is seen that individual terms in eqs.~(\ref{eq:V20})-(\ref{eq:V23})
 can be unambiguously
identified with the pertinent terms in the
multiple-scattering series that emerges from the system
of equations displayed in section~\ref{sec:faddeev}. 
Moreover,  as already mentioned before, it is seen that,
in order to make these two series coincide, one should omit the terms of the
type \textit{$\btheta_\alpha{\bf G}_{\sf F}\btheta_\beta$ with $\alpha\neq\beta$}
in the equation for \textit{${\bf R}_{\alpha\beta}$ }(third line in eq.~(\ref{eq:R})).

The triple scattering is relegated to Appendix~\ref{app:triple}. It can be 
checked again that
to this order the multiple-scattering series which are produced by the system of equations 
displayed in section~\ref{sec:faddeev} can be unambiguously identified. The generalization
to the higher orders is straightforward.

\subsection{Induced three-body force} 

The inclusion of the induced three-body force, which is described by the operator \textit{${\bf K}_4$},
 does not lead to any complication, since \textit{${\bf K}_4$} is a smooth function of all momenta.
Moreover, at this stage it is seen that a genuine three-body force could be included
without any further ado.

\subsection{Analytic continuation and the cusp effect}

In the three-body counterpart of the L\"uscher formula, the 
{\em on-shell} \textit{$T$}-matrix elements {\em below threshold} are necessarily present, due
to the fact that the on-shell projector \textit{$\Delta$} defined in 
eq.~(\ref{eq:testfunction}) does not vanish for \textit{$q_0^2<0$}. Instead, a smooth
regulator \textit{$f(q_0^2/\mu^2)$} is introduced
in eq.~(\ref{eq:testfunction}), which effectively suppresses the contributions
with \textit{$-q_0^2>\mu^2$}. However, since in the original sum the variable \textit{${\bf p}^2$}
was positively defined, a question naturally arises, whether such an analytic
continuation is needed at all. Indeed, as it is known, one can avoid this 
procedure in the two-particle case, even if multiple scattering channels are
considered.

The reason why the analytic continuation in the three-particle case can not
be avoided, lies in the following: if, instead of the smooth function
\textit{$f$}, a sharp cutoff \textit{$\theta(q_0^2)$} is introduced, the principal-value integral
will have {\em a unitary cusp} at threshold, proportional
to the factor \textit{$\sqrt{-q_0^2}$}. In the two-particle case, the quantity 
\textit{$q_0^2$} depends on \textit{$z$} only and the cusp does not cause a problem.
In the three-particle case, \textit{$q_0^2$} depends in addition on the 
spectator momentum \textit{${\bf p}_\alpha$}, see eq.~(\ref{eq:q02a}). Due to the presence
of the cusp, one can not apply the regular summation theorem: the finite
volume corrections are suppressed by a power of \textit{$L$}, not by exponentials.
In order to see this, let us consider a simple example with the cusp present
\textit{\eq
I=\dfrac{1}{L^3}\,\sum_{\bf p}|{\bf p}|\,\theta(\Lambda^2-{\bf p}^2)
=\int^\Lambda d^3{\bf k}\,|{\bf k}|\,\dfrac{1}{L^3}\,\sum_{\bf p}
\delta^3({\bf p}-{\bf k})\, .
\en }
Using Poisson's summation formula, the above expression can be transformed into
\textit{\eq\label{eq:L-2}
I=\int^\Lambda \dfrac{d^3{\bf k}}{(2\pi)^3}\,|{\bf k}|
+\sum_{{\bf n}\in \mathbb{Z}^3\backslash{\bf 0}}
\int^\Lambda \dfrac{d^3{\bf k}}{(2\pi)^3}\,|{\bf k}|\,e^{i{\bf n}{\bf k}L}=I_1+I_2\, .
\en }
Carrying out the integration in the second term, it is easy to see 
that it vanishes as \textit{$L^{-2}$} and not as an exponential, see 
Appendix~\ref{app:L-2}.

To summarize, the price to pay for the exponential fall-off of the finite-volume
corrections in the regular terms\footnote{More precisely, the finite-volume 
corrections in the regular terms fall off faster than any inverse power of 
\textit{$L$}~\cite{luescher-2.}.} that contain principal-value integrals is
that these principal-value integrals should be defined through the analytic
continuation below threshold. If one wishes to use the information from the
physical region only, the finite-volume corrections in the regular terms
will be power-suppressed, not exponentially suppressed.

\section{Conclusions}
\label{sec:concl}

\begin{itemize}

\item[i)]
In this paper we have derived the three-body counterpart of the L\"uscher formula.
The pertinent
expressions are displayed in section~\ref{sec:faddeev}.

\item[ii)]
The fundamental property of the finite-volume spectrum, which follows from
this formula, is that the spectrum is completely determined
by the \textit{$S$}-matrix elements for the transitions \textit{$2\to 2$},\textit{$2\to 3$} and \textit{$3\to 3$}
in the infinite volume. Consequently, two different potential models
with the same \textit{$S$}-matrix elements lead to the same spectra up 
to the exponentially
suppressed corrections.

\item[iii)]
The equations given in section~\ref{sec:faddeev} have a complicated structure.
However, due to the presence of the on-shell 
factor \textit{$\Delta({\bf k}^2,q_0^2)$}, the
dimensionality of equations is reduced as compared to the original Faddeev
equations. Namely, in the Faddeev equations we have two momenta describing the
three-particle intermediate state. The integration over one of these momenta is
removed by the on-shell factor. This is similar to the conventional L\"uscher
formula, which becomes an algebraic equation 
(after the truncation of the partial-wave expansion), whereas the original
Lippmann-Schwinger equation was an integral equation in one momentum variable.
Despite this simplification, 
a direct numerical solution of the three-body equations in a finite volume,
as described in section~\ref{sec:model-finite}, may still prove to be less
challenging.

\item[iv)]
In this paper, we restrict ourselves to the non-relativistic potential model.
Considering the processes within the field theory will introduce several novel aspects,
\textit{e.g.}, relativistic effects, particle creation and annihilation, etc.
However, we do not expect that these effects will upset the proof
given in the present paper. Further investigations are planned in this
direction in the future, and the results will be reported elsewhere.

\item[v)]
It is legitimate to ask, whether it is possible 
to use equations from section~\ref{sec:faddeev}
in order to extract the \textit{$S$}-matrix elements from 
the measured spectrum. Due to the
complex nature of these equations, a straightforward extraction can 
be very complicated. Adopting a strategy similar to that
of ref.~\cite{oset} -- introducing a phenomenologically 
reasonable parameterization
of the potential and fitting the parameters of the potential to the 
lattice data -- could be more promising.
The physical observables in the infinite volume 
(\textit{$S$}-matrix elements, resonance pole positions)
can be then obtained from the solution of the scattering equations.

\item[vi)]
From the above discussion it becomes clear that it would be very interesting to
carry out calculations of the finite-volume spectrum in different 
realistic models
(see, \textit{e.g.}~\cite{Doring:2009yv1,Doring:2009yv2}),
which predict both the Roper resonance and $N(1535)$ in the infinite volume.
These calculations may shed light on the puzzle related to the ``wrong''
level ordering for such systems in lattice QCD.
The findings of the present paper guarantee that the spectrum will stay
stable with respect to the choice of a model and the variation 
of its parameters, provided the
\textit{$S$}-matrix elements in the infinite volume remain the same.

\end{itemize}

\section*{Acknowledgments}

\begin{sloppypar}
The authors would like to thank A. Anisovich, S. Beane, M. D\"oring, J. Gasser, 
D. Lee, U.-G. Mei{\ss}ner, H. Meyer, E. Oset, A. Sarantsev and M. Savage
for interesting discussions.
This work is partly supported by the EU
Integrated Infrastructure Initiative HadronPhysics3 Project 
We also acknowledge the support by DFG (SFB/TR 16,
``Subnuclear Structure of Matter'') 
and by COSY FFE under contract 41821485 (COSY 106). 
A.R. acknowledges  support of the Georgia National Science Foundation (Grant
\#GNSF/ST08/4-401).
\end{sloppypar}

\renewcommand{\thefigure}{\thesection.\arabic{figure}}
\renewcommand{\thetable}{\thesection.\arabic{table}}
\renewcommand{\theequation}{\thesection.\arabic{equation}}

\appendix
\setcounter{figure}{0}
\setcounter{table}{0}
\setcounter{equation}{0}
\section{Three identical particles with a separable potential}
\label{app:separable}

In this Appendix, we list the expressions, which appeared first 
in section~\ref{sec:model}, in a special case of the pair 
potentials having  separable form. Further, we
restrict ourselves to the case
of three identical particles with the masses \textit{$m_1=m_2=m_3=m$}.

The free-particle states of \textit{$n$} identical particles are normalized, according to
\textit{\eq
|{\bf k}_1\cdots {\bf k}_n\rangle=\dfrac{1}{\sqrt{n!}}\,
a^\dagger({\bf k}_1)\cdots a^\dagger({\bf k}_n)|0\rangle\, .
\en }
In case of the identical particles, there is only one term in the two-body
Hamiltonian \textit{${\bf H}_{\sf 2\to 2}$}, given by eq.~(\ref{eq:H}) -- 
the sum over \textit{$\alpha,\beta$} drops out. Introducing the appropriate symmetry factors,
the two-particle potential in the separable model can be written in the form
\textit{\eq
\bar V_\alpha({\bf q}',{\bf q})=\dfrac{1}{2!}\,v({\bf q}')v({\bf q})\, .
\en }
Assuming that the vertex \textit{$\bar\Gamma_\alpha$} is also separable
\textit{\eq
\bar\Gamma_\alpha({\bf q}';{\bf p},{\bf q})=\dfrac{1}{\sqrt{2!3!}}\,
\lambda({\bf q}')f({\bf p},{\bf q})\, .
\en }
The effective two-particle potential then can be written as
\textit{\eq
\langle {\bf q}'|{\bf w}(z)|{\bf q}\rangle
=-v({\bf q}')v({\bf q})+\lambda({\bf q}')d(z)\lambda({\bf q})\, ,
\en }
where \textit{$d(z)$} is given by
\textit{\eq\label{eq:dz}
&&\!d(z)=\int
\dfrac{d^3{\bf p}}{(2\pi)^3}\,\dfrac{d^3{\bf q}}{(2\pi)^3}\,
\dfrac{f^2({\bf p},{\bf q})}{3m+\dfrac{3{\bf p}^2}{4m}+\dfrac{{\bf q}^2}{m}-z-i0}
\nonumber\\[2mm]
&+&\!\int
\dfrac{d^3{\bf p}'}{(2\pi)^3}\,\dfrac{d^3{\bf q}'}{(2\pi)^3}
\dfrac{d^3{\bf p}}{(2\pi)^3}\,\dfrac{d^3{\bf q}}{(2\pi)^3}\,
\dfrac{f({\bf p}',{\bf q}')
\langle{\bf p}'{\bf q}'|{\bf \hat m}(z)|{\bf p}{\bf q}\rangle
f({\bf p},{\bf q})}
{\biggl(3m+\dfrac{3({\bf p}')^2}{4m}+\dfrac{({\bf q}')^2}{m}-z-i0\biggr)
\biggl(3m+\dfrac{3{\bf p}^2}{4m}+\dfrac{{\bf q}^2}{m}-z-i0\biggr)}\, ,
\nonumber\\
&&
\en }
where
\textit{\eq
\langle{\bf p}'{\bf q}'|{\bf \hat m}(z)|{\bf p}{\bf q}\rangle
&=&
\dfrac{1}{3}\,\sum_{\alpha,\beta=1}^3v({\bf q}'_\alpha)
\biggl\{(2\pi)^3\delta^3({\bf p}'_\alpha-{\bf p}_\beta)
\tau\biggl(z-m-\dfrac{3{\bf p}_\alpha^2}{4m}\biggr)
\nonumber\\[2mm]
&+&
\tau\biggl(z-m-\dfrac{3({\bf p}'_\alpha)^2}{4m}\biggr)
\langle{\bf p}'_\alpha|{\bf Y}(z)|{\bf p}_\beta\rangle
\tau\biggl(z-m-\dfrac{3{\bf p}_\beta^2}{4m}\biggr)\biggr\}v({\bf q}_\beta)\, ,
\en }
with
\textit{\eq
{\bf p}_1=-\dfrac{1}{2}\,{\bf p}+{\bf q}\, ,\quad
{\bf p}_2=-\dfrac{1}{2}\,{\bf p}-{\bf q}\, ,\quad
{\bf p}_3={\bf p}\, ,
\nonumber\\[2mm]
{\bf q}_1=-\dfrac{3}{4}\,{\bf p}-\dfrac{1}{2}\,{\bf q}\, ,\quad
{\bf q}_2=\dfrac{3}{4}\,{\bf p}-\dfrac{1}{2}\,{\bf q}\, ,\quad
{\bf q}_3={\bf q}\, ,
\en }
and similarly for \textit{${\bf p}'_\alpha,{\bf q}'_\alpha$}.

Further, the quantity \textit{$\tau(z)$} is defined by
\textit{\eq\label{eq:tau}
\tau(z)=-\biggr(1+\int\dfrac{d^3{\bf q}}{(2\pi)^3}\,
\dfrac{v^2({\bf q})}{2m+\dfrac{{\bf q}^2}{m}-z-i0}\biggr)^{-1}\, .
\en }
Finally, the quantity \textit{$\langle{\bf p}'|{\bf Y}(z)|{\bf p}\rangle$} obeys the equation
\textit{\eq\label{eq:Y}
\langle{\bf p}'|{\bf Y}(z)|{\bf p}\rangle
=2Z({\bf p}',{\bf p})+\int\dfrac{d^3{\bf p}''}{(2\pi)^3}\,
2Z({\bf p}',{\bf p}'')\tau\biggl(z-m-\dfrac{3({\bf p}'')^2}{4m}\biggr)
\langle{\bf p}''|{\bf Y}(z)|{\bf p}\rangle\, ,
\en }
with the kernel
\textit{\eq
Z({\bf p}',{\bf p})=\dfrac{v({\bf p}'/2+{\bf p})
v({\bf p}'+{\bf p}/2)}
{3m+\dfrac{({\bf p}')^2+{\bf p}^2+{\bf p}'{\bf p}}{m}-z-i0}\, .
\en }

\setcounter{figure}{0}
\setcounter{table}{0}
\setcounter{equation}{0}
\section{Separable potential in a finite volume}
\label{app:separableL}

Below, as in Appendix~\ref{app:separable},
we restrict ourselves to the case of three identical particles interacting with
the separable pair potentials. The finite-volume versions of 
eqs.~(\ref{eq:dz})-(\ref{eq:Y}) are obtained by merely
replacing the integrals by the sums.
We do not display these equations explicitly, with one exception:
the finite-volume version of eq.~(\ref{eq:tau}) is given by 
\textit{\eq\label{eq:tauL}
\tau^L(z)=-\biggr(1+\dfrac{1}{L^3}\sum_{\bf q}
\dfrac{v^2({\bf q})}{2m+\dfrac{{\bf q}^2}{m}-z}\biggr)^{-1}\, ,
\en }
where, as already mentioned above, the sum over the momentum \textit{${\bf q}$} runs over
(cf. eq.~(\ref{eq:JacobiL}))
\textit{\eq
{\bf q}=\dfrac{2\pi}{L}\biggl({\bf l}+\dfrac{1}{2}\,{\bf n}\biggr)\, ,\quad\quad
{\bf l},{\bf n}=\mathbb{Z}^3\, .
\en }
Here, the shift of the discrete variable \textit{${\bf l}$} is related to the CM motion
of a two-particle pair in the rest frame of three particles. 

Using the regular summation theorem~\cite{luescher-2.}, we obtain that, up to the
terms exponentially suppressed at large values of \textit{$L$},
\textit{\eq
&&\tau^L(z(p))=\biggl(\dfrac{mv^2(p)}{4\pi}\biggr)^{-1}
\biggl(p\cot\delta(p)-\dfrac{2}{\sqrt{\pi}L}Z_{00}^\theta(1;\nu^2)\biggr)^{-1}\, ,
\quad z(p)=2m+\dfrac{p^2}{m}\, ,
\nonumber\\[2mm]
&&p\cot\delta(p)=-\dfrac{4\pi}{mv^2(p)}\,\biggl(1+\dfrac{m}{2\pi^2}\,{\sf P.V.}\int_0^\infty\dfrac{dq\, q^2 v^2(q)}{q^2-p^2}\biggr)\, ,
\nonumber\\[2mm]
&&Z_{00}^\theta(1;\nu^2)=\dfrac{1}{\sqrt{4\pi}}\,\sum_{{\bf l}\in\mathbb{Z}^3}
\dfrac{1}{({\bf l}+\btheta/2\pi)^2-\nu^2}\, ,\quad \nu=\dfrac{pL}{2\pi}\, ,\quad
\theta_i=2\pi\biggl(\dfrac{n_i}{2}-\biggl[\dfrac{n_i}{2}\biggr]\biggr)\, .
\en }
In other words, the components \textit{$\theta_i$} of the vector \textit{$\btheta$} are either \textit{$0$} or \textit{$\pi$},
for the even/odd components of the vector \textit{${\bf n}$}.

It is interesting to compare the above equations to the three-body equations in a finite
volume, which were considered in refs.~\cite{Kreuzer:2008bi,Kreuzer:2009jp,Kreuzer:2010ti}.
A straightforward comparison leads to the conclusion that the sole difference between these
equations consists in the replacement \textit{$Z_{00}^\theta(1;\nu^2)$} by \textit{$Z_{00}(1;\nu^2)$}, \textit{i.e.},
in neglecting the CM motion of the two-particle sub-systems in the rest frame of
three particles. Note also that the parameter \textit{$\btheta$} is related to the topological phase
considered in ref.~\cite{Bour:2011ef}.

\setcounter{figure}{0}
\setcounter{table}{0}
\setcounter{equation}{0}
\section{Summation in eq.~(\ref{eq:summation})}
\label{app:summation}

The multiple-scattering series in eq.~(\ref{eq:appendix}) can be reproduced 
through the following system of linear equations
\textit{\eq\label{eq:matrix}
\begin{pmatrix}T_1\cr T_2\cr T_3\end{pmatrix}=
\begin{pmatrix}1\cr 1\cr 1\end{pmatrix}+
\begin{pmatrix}0 & y_2 & y_3\cr y_1 & 0 & y_3\cr y_1 & y_2 & 0\end{pmatrix}
\begin{pmatrix}T_1\cr T_2\cr T_3\end{pmatrix}\, .
\en }
It is straightforward to ensure that
\textit{\eq
\!&&
\dfrac{1}{2}\,(T_1+T_2+T_3-1)=1+(y_1+y_2+y_3)+(y_1(y_2+y_3)+y_2(y_1+y_3)+y_3(y_1+y_2))+\cdots\, ,
\nonumber\\
\!&&
\en }
which has exactly the same structure as the series in  
eq.~(\ref{eq:appendix}).

On the other hand, eq.~(\ref{eq:matrix}) can be solved directly. 
The solution of this equation is given in eq.~(\ref{eq:summation}).

\setcounter{figure}{0}
\setcounter{table}{0}
\setcounter{equation}{0}
\section{The asymptotic behavior of the last term in eq.~(\ref{eq:L-2})}
\label{app:L-2}

Carrying out the integration in \textit{${\bf k}$}, the last term in eq.~(\ref{eq:L-2})
can be written as
\textit{\eq\label{eq:I2}
I_2=\dfrac{1}{2\pi^2L^4}\sum_{{\bf n}\in\mathbb{Z}\backslash{\bf 0}}
\biggl\{-\dfrac{2}{|{\bf n}|^4}+2\,\mbox{{\rm Re}}\,J_4(x)+2x\,\mbox{ {\rm Im}}\,J_3(x)
-x^2\,\mbox{{\rm Re}}\,J_2(x)\biggr\}\, ,\quad\quad x=L\Lambda\, ,
\en }
where
\textit{\eq\label{eq:Ji}
J_i(z)=\sum_{{\bf n}\in\mathbb{Z}\backslash{\bf 0}}\dfrac{z^{|{\bf n}|}}{|{\bf n}|^i}\, ,
\quad\quad
z=e^{ix}\, .
\en }
Using the relation from ref.~\cite{Doring:2011ip}
\textit{\eq
\sum_{{\bf n}\in\mathbb{Z}}z^{|{\bf n}|}=g_P(z)\, \quad\quad
g_P(z)=(\theta_3(0,z))^3\, ,
\en }
where \textit{$\theta_3(0,z)$} is the elliptic \textit{$\theta$}-function
\textit{\eq\label{eq:theta-sum}
\theta_3(0,z)=\sum_{k=-\infty}^{\infty}z^{k^2}\, ,
\en }
which obeys the following integral representation,
\textit{\eq\label{eq:theta-integral}
\theta_3(0,z)=-i\int_{i-\infty}^{i+\infty}du\,z^{u^2}\cot(\pi u)\, ,
\en }
a set of differential equations, which relate \textit{$J_i(z)$} with \textit{$g_P(z)$}, can be derived. For example, for \textit{$i=1$}, the pertinent equation has the form
\textit{\eq
\dfrac{dJ_1(z)}{dz}=\dfrac{g_P(z)-1}{z}\, .
\en }
Equations for \textit{$i=2,3,4,\cdots $} can be obtained in the similar fashion.
Integrating these differential equations, 
the following expression for \textit{$J_i(z)$} can be straightforwardly obtained
\textit{\eq
J_i(z)=\dfrac{(-1)^{i-1}}{(i-1)!}\,\int_0^1 dy\, (\ln y)^{i-1}\dfrac{g_P(zy)-1}{y}\, ,
\quad\quad
i=2,3,4\, .
\en }
Further, from eq.~(\ref{eq:theta-sum}) it follows that
\textit{\eq
|\theta_3(0,ye^{ix})|\leq \theta_3(0,y)\, ,\quad\quad
 \theta_3(0,y)\geq 1\, ,\quad\quad\mbox{for}~y\geq 0\, .
\en }
From the above equation, we readily obtain
\textit{\eq
|(\theta_3(0,ye^{ix}))^3-1|\leq (\theta_3(0,y))^3-1\, .
\en } 
Consequently,
\textit{\eq
|J_i(z)|\leq
\dfrac{1}{(i-1)!}\,\int_0^1dy|\ln y|^{i-1}\dfrac{|g_P(y)-1|}{y}\, .
\en }
The above integral converges at \textit{$y=0$}. Further, since the series in
 eq.~(\ref{eq:theta-sum}) converges for \textit{$|y|<1$}, the divergence
in the integral may occur only on the upper limit \textit{$y=1$}. Indeed, using
the integral representation, one finds that 
 \textit{$\theta_3(0,y)\sim (1-y)^{-1/2}$} as \textit{$y\to 1^-$}. However, due to the fact that
\textit{$\ln y$} vanishes as \textit{$y\to 1$}, the singularity in the
integrand is of the integrable type. Consequently, \textit{$|J_i(z)|$}are uniformly
bound from above and therefore, the quantity \textit{$L^2I_2$} is also bound at
 \textit{$L\to\infty$}. 

Last but not least, we wish to address here the issue of using a sharp cutoff
at a momentum \textit{$\Lambda$}. As seen, \textit{e.g.}, from eqs.~(\ref{eq:I2}) and 
(\ref{eq:Ji}), the expression for the quantity \textit{$I_2$} contains a sum of 
rapidly oscillating terms proportional to \textit{$\sin(|{\bf n}|L\Lambda)$} and
\textit{$\cos(|{\bf n}|L\Lambda)$}, as \textit{$L\to\infty$}. These terms are the artifacts
of using a sharp cutoff and should disappear when the cutoff is removed, since
no observable effect in the infrared should emerge from the 
ultraviolet cutoff. However, the expressions diverge at \textit{$\Lambda\to\infty$}.
In order to tackle this problem, the simplest way is to introduce an
additional smooth cutoff in the expressions
\textit{\eq
I_2=\sum_{{\bf n}\in\mathbb{Z}\backslash{\bf 0}} J_n(\varepsilon)\, ,\quad\quad
  J_n(\varepsilon)=\int^\Lambda\dfrac{d^3{\bf p}}{(2\pi)^3}\,
\exp(-\varepsilon{\bf p}^2)\,|{\bf p}|e^{i{\bf n}{\bf p}L}\, ,
\en }
and then consider the limits \textit{$\Lambda\to\infty$} and \textit{$\varepsilon\to 0$}
(in this order). It is easy to show that
\textit{\eq
\lim_{\varepsilon\to 0}\lim_{\Lambda\to\infty}J_n=\dfrac{1}{2\pi^2L^4}\,
\dfrac{1}{{\bf n}^4}\,\lim_{\varepsilon'\to 0}\int_0^\infty dp\,p^2
e^{-\varepsilon'p^2}\sin p\, ,
\quad\quad \varepsilon'=\dfrac{\varepsilon}{{\bf n}^2L^2}\, .
\en }
Now using the equality~\cite{Gradsteyn}
\textit{\eq
\lim_{\varepsilon'\to 0}\int_0^\infty dp\, p^2
e^{-\varepsilon'p^2}\sin p=\lim_{\varepsilon'\to 0}\biggl(-\dfrac{d}{d\varepsilon'}
\biggl\{\dfrac{1}{2\varepsilon'}\!~_1F_1\biggl(1;\dfrac{3}{2};-\dfrac{1}{4\varepsilon'}\biggr)\biggr\}\biggr)=-2\, ,
\en }
where \textit{$\!~_1F_1(a;b;z)$} denotes the confluent hypergeometric function, we finally get
\textit{\eq
I_2=-\dfrac{1}{\pi^2L^4}\sum_{{\bf n}\in\mathbb{Z}\backslash{\bf 0}}
\dfrac{1}{|{\bf n}|^4}=O(L^{-4})\, .
\en }
Comparing with eq.~(\ref{eq:I2}), one sees that all the oscillating terms
disappear, as expected. However, the final expression is still only power-suppressed in \textit{$L$}, not exponentially suppressed.

\setcounter{figure}{0}
\setcounter{table}{0}
\setcounter{equation}{0}
\section{Triple scattering diagram}
\label{app:triple}

Below we shall consider the triple scattering diagram, which is given by
\textit{\eq
&&I_3^c=\sum_{\alpha\neq\beta,\beta\neq\gamma}
\dfrac{1}{L^{15}}\,\sum_{{\bf p}''_\alpha{\bf p}'_\beta{\bf p}_\gamma{\bf q}''_\alpha{\bf q}_\gamma}
\dfrac{\tilde\Gamma_\alpha({\bf q}';{\bf p}''_\alpha{\bf q}''_\alpha)
(-\bar V_\alpha({\bf q}''_\alpha,{\bf p}'_\beta+\dfrac{m_\beta}{M-m_\alpha}\,
{\bf p}''_\alpha))}
{\biggl(M+\dfrac{({\bf p}''_\alpha)^2}{2M_\alpha}
+\dfrac{({\bf q}''_\alpha)^2}{2\mu_\alpha}-z\biggr)}
\nonumber\\[2mm]
&\times&\dfrac{(-\bar V_\beta(-{\bf p}''_\alpha- \dfrac{m_\alpha}{M-m_\beta}\,
{\bf p}'_\beta,-{\bf p}_\gamma-\dfrac{m_\gamma}{M-m_\beta}\,{\bf p}'_\beta))
(-\bar V_\gamma({\bf p}'_\beta+\dfrac{m_\beta}{M-m_\gamma}\,{\bf p}_\gamma,
{\bf q}_\gamma))}
{\biggl(M+\dfrac{({\bf p}'_\beta)^2}{2M_\beta}
+\dfrac{(-{\bf p}''_\alpha-\dfrac{m_\alpha}{M-m_\beta}\,
{\bf p}'_\beta)^2}{2\mu_\beta}-z\biggr)}
\nonumber\\[2mm]
&\times&\dfrac{(\tilde\Gamma_\gamma({\bf q};{\bf p}_\gamma{\bf q}_\gamma))^*}
{\biggl(M+\dfrac{{\bf p}_\gamma^2}{2M_\gamma}
+\dfrac{{\bf q}_\gamma^2}{2\mu_\gamma}-z\biggr)
\biggl(M+\dfrac{({\bf p}'_\beta)^2}{2M_\beta}
+\dfrac{(-{\bf p}_\gamma-\dfrac{m_\gamma}{M-m_\beta}\,
{\bf p}'_\beta)^2}{2\mu_\beta}-z\biggr)}\, .
\en }
In this expression, first the summations over the momenta 
\textit{${\bf q}_\alpha'',{\bf q}_\gamma$}
are carried out, similarly as in section~\ref{sec:connected}. Further,
with the use of the formal relation (see Appendix~\ref{app:relation})
\textit{\eq\label{eq:relation}
&&\hspace*{-.5cm}\dfrac{1}{L^3}\sum_{{\bf k}'_\beta}\dfrac{(2\pi)^3\delta^3({\bf p}'_\beta-{\bf k}'_\beta)}
{\biggl(M+\dfrac{({\bf p}''_\alpha)^2}{2M_\alpha}
+\dfrac{({\bf k}'_\beta+\dfrac{m_\beta}{M-m_\alpha}\,
{\bf p}''_\alpha)^2}{2\mu_\alpha}-z\biggr)
\biggl(M+\dfrac{({\bf p}_\gamma)^2}{2M_\gamma}
+\dfrac{({\bf k}'_\beta+\dfrac{m_\beta}{M-m_\gamma}\,
{\bf p}_\gamma)^2}{2\mu_\gamma}-z\biggr)}
\nonumber\\[2mm]
&&\hspace*{2.5cm}=\,\,\,G_{{\sf K}\alpha}\biggl({\bf p}''_\alpha,{\bf p}'_\beta+\dfrac{m_\beta}{M-m_\alpha}\,
{\bf p}''_\alpha;z\biggr)
G_{{\sf K}\gamma}\biggl({\bf p}_\gamma,{\bf p}'_\beta+\dfrac{m_\beta}{M-m_\gamma}\,
{\bf p}_\gamma;z\biggr)
\nonumber\\[2mm]
&&\hspace*{2.5cm}+\,\,\,\dfrac{\tilde G_{{\sf F}\alpha}\biggl({\bf p}''_\alpha,{\bf p}'_\beta+\dfrac{m_\beta}{M-m_\alpha}\,
{\bf p}''_\alpha;z\biggr)}
{\biggl(M+\dfrac{({\bf p}_\gamma)^2}{2M_\gamma}
+\dfrac{({\bf p}'_\beta+\dfrac{m_\beta}{M-m_\gamma}\,
{\bf p}_\gamma)^2}{2\mu_\gamma}-z\biggr)}
\nonumber\\[2mm]
&&\hspace*{2.5cm}+\,\,\,\dfrac{\tilde G_{{\sf F}\gamma}\biggl({\bf p}_\gamma,{\bf p}'_\beta+\dfrac{m_\beta}{M-m_\gamma}\,
{\bf p}_\gamma;z\biggr)}
{\biggl(M+\dfrac{({\bf p}''_\alpha)^2}{2M_\alpha}
+\dfrac{({\bf p}'_\beta+\dfrac{m_\beta}{M-m_\alpha}\,
{\bf p}''_\alpha)^2}{2\mu_\alpha}-z\biggr)}\, ,
\en }
as well as the symmetry properties of the denominators
\textit{\eq
\dfrac{({\bf p}_\beta')^2}{2M_\beta}
+\dfrac{({\bf p}_\alpha''+\dfrac{m_\alpha}{M-m_\beta}\,{\bf p}_\beta')^2}{2\mu_\beta}
&=&\dfrac{({\bf p}''_\alpha)^2}{2M_\alpha}
+\dfrac{({\bf p}'_\beta+\dfrac{m_\beta}{M-m_\alpha}\,
{\bf p}''_\alpha)^2}{2\mu_\alpha}
\nonumber\\[2mm]
\dfrac{({\bf p}_\beta')^2}{2M_\beta}
+\dfrac{({\bf p}_\gamma+\dfrac{m_\gamma}{M-m_\beta}\,{\bf p}_\beta')^2}{2\mu_\beta}
&=&\dfrac{({\bf p}_\gamma)^2}{2M_\gamma}
+\dfrac{({\bf p}'_\beta+\dfrac{m_\beta}{M-m_\gamma}\,
{\bf p}_\gamma)^2}{2\mu_\gamma}\, ,
\en }
the above expression can be rewritten in the following form
\textit{\eq
I_2^c&=&\sum_{\alpha\neq\beta,\beta\neq\gamma}
\int
\dfrac{d^3{\bf q}''_\alpha}{(2\pi)^3}
\dfrac{d^3{\bf q}_\gamma}{(2\pi)^3}\,
\biggl\{
\nonumber\\[2mm]
&\times&
\int
\dfrac{d^3{\bf p}''_\alpha}{(2\pi)^3}
\dfrac{d^3{\bf p}'_\beta}{(2\pi)^3}
\dfrac{d^3{\bf p}_\gamma}{(2\pi)^3}
\tilde\Gamma_\alpha({\bf q}';{\bf p}''_\alpha{\bf q}''_\alpha)
U^{c,0}_{3\alpha\beta\gamma}({\bf p}''_\alpha{\bf q}''_\alpha;{\bf p}'_\beta;
{\bf p}_\gamma{\bf q}_\gamma)
(\tilde\Gamma_\gamma({\bf q};{\bf p}_\gamma{\bf q}_\gamma))^*
\nonumber\\[2mm]
&+&\dfrac{1}{L^3}\sum_{{\bf p}''_\alpha}
\int
\dfrac{d^3{\bf p}'_\beta}{(2\pi)^3}
\dfrac{d^3{\bf p}_\gamma}{(2\pi)^3}
\tilde\Gamma_\alpha({\bf q}';{\bf p}''_\alpha{\bf q}''_\alpha)
U^{c,1}_{3\alpha\beta\gamma}({\bf p}''_\alpha{\bf q}''_\alpha;{\bf p}'_\beta;
{\bf p}_\gamma{\bf q}_\gamma)
(\tilde\Gamma_\gamma({\bf q};{\bf p}_\gamma{\bf q}_\gamma))^*
\nonumber\\[2mm]
&+&\dfrac{1}{L^3}\sum_{{\bf p}'_\beta}
\int
\dfrac{d^3{\bf p}''_\alpha}{(2\pi)^3}
\dfrac{d^3{\bf p}_\gamma}{(2\pi)^3}
\tilde\Gamma_\alpha({\bf q}';{\bf p}''_\alpha{\bf q}''_\alpha)
U^{c,2}_{3\alpha\beta\gamma}({\bf p}''_\alpha{\bf q}''_\alpha;{\bf p}'_\beta;
{\bf p}_\gamma{\bf q}_\gamma)
(\tilde\Gamma_\gamma({\bf q};{\bf p}_\gamma{\bf q}_\gamma))^*
\nonumber\\[2mm]
&+&\dfrac{1}{L^3}\sum_{{\bf p}_\gamma}
\int
\dfrac{d^3{\bf p}''_\alpha}{(2\pi)^3}
\dfrac{d^3{\bf p}'_\beta}{(2\pi)^3}
\tilde\Gamma_\alpha({\bf q}';{\bf p}''_\alpha{\bf q}''_\alpha)
U^{c,3}_{3\alpha\beta\gamma}({\bf p}''_\alpha{\bf q}''_\alpha;{\bf p}'_\beta;
{\bf p}_\gamma{\bf q}_\gamma)
(\tilde\Gamma_\gamma({\bf q};{\bf p}_\gamma{\bf q}_\gamma))^*
\nonumber\\[2mm]
&+&\dfrac{1}{L^6}\sum_{{\bf p}'_\beta{\bf p}_\gamma}
\int
\dfrac{d^3{\bf p}''_\alpha}{(2\pi)^3}
\tilde\Gamma_\alpha({\bf q}';{\bf p}''_\alpha{\bf q}''_\alpha)
U^{c,4}_{3\alpha\beta\gamma}({\bf p}''_\alpha{\bf q}''_\alpha;{\bf p}'_\beta;
{\bf p}_\gamma{\bf q}_\gamma)
(\tilde\Gamma_\gamma({\bf q};{\bf p}_\gamma{\bf q}_\gamma))^*
\nonumber\\[2mm]
&+&\dfrac{1}{L^6}\sum_{{\bf p}''_\alpha{\bf p}'_\beta}
\int
\dfrac{d^3{\bf p}_\gamma}{(2\pi)^3}
\tilde\Gamma_\alpha({\bf q}';{\bf p}''_\alpha{\bf q}''_\alpha)
U^{c,5}_{3\alpha\beta\gamma}({\bf p}''_\alpha{\bf q}''_\alpha;{\bf p}'_\beta;
{\bf p}_\gamma{\bf q}_\gamma)
(\tilde\Gamma_\gamma({\bf q};{\bf p}_\gamma{\bf q}_\gamma))^*
\nonumber\\[2mm]
&+&\dfrac{1}{L^6}\sum_{{\bf p}''_\alpha{\bf p}_\gamma}
\int
\dfrac{d^3{\bf p}'_\beta}{(2\pi)^3}
\tilde\Gamma_\alpha({\bf q}';{\bf p}''_\alpha{\bf q}''_\alpha)
U^{c,6}_{3\alpha\beta\gamma}({\bf p}''_\alpha{\bf q}''_\alpha;{\bf p}'_\beta;
{\bf p}_\gamma{\bf q}_\gamma)
(\tilde\Gamma_\gamma({\bf q};{\bf p}_\gamma{\bf q}_\gamma))^*\biggr\}\, ,
\en }
where
\textit{\eq
U^{c,0}_{3\alpha\beta\gamma}({\bf p}''_\alpha{\bf q}''_\alpha;{\bf p}'_\beta;
{\bf p}_\gamma{\bf q}_\gamma)
&=&G_{{\sf K}\alpha}({\bf p}''_\alpha,{\bf q}''_\alpha;z)
(-\bar V_\alpha({\bf q}''_\alpha,\bar {\bf q}''_\alpha))
G_{{\sf K}\alpha}({\bf p}''_\alpha,\bar {\bf q}''_\alpha;z)
(-\bar V_\beta(\bar {\bf q}''_\beta,\bar {\bf q}'''_\beta))
\nonumber\\[2mm]
&\times&
G_{{\sf K}\beta}({\bf p}'_\beta,\bar {\bf q}'''_\beta;z)
(-\bar V_\gamma(\bar {\bf q}''_\gamma,{\bf q}_\gamma))
G_{{\sf K}\gamma}({\bf p}_\gamma,{\bf q}_\gamma;z)
\en }
\textit{\eq
U^{c,1}_{3\alpha\beta\gamma}({\bf p}''_\alpha{\bf q}''_\alpha;{\bf p}'_\beta;
{\bf p}_\gamma{\bf q}_\gamma)
&=&\tilde G_{{\sf F}\alpha}({\bf p}''_\alpha,{\bf q}''_\alpha;z)
(-\bar V_\alpha({\bf q}''_\alpha,\bar {\bf q}''_\alpha))
G_{{\sf K}\alpha}({\bf p}''_\alpha,\bar {\bf q}''_\alpha;z)
(-\bar V_\beta(\bar {\bf q}''_\beta,\bar {\bf q}'''_\beta))
\nonumber\\[2mm]
&\times&
G_{{\sf K}\beta}({\bf p}'_\beta,\bar {\bf q}'''_\beta;z)
(-\bar V_\gamma(\bar {\bf q}''_\gamma,{\bf q}_\gamma))
G_{{\sf K}\gamma}({\bf p}_\gamma,{\bf q}_\gamma;z)
\nonumber\\[2mm]
&+&\tilde G_{{\sf F}\alpha}({\bf p}''_\alpha,{\bf q}''_\alpha;z)
(-\bar V_\alpha({\bf q}''_\alpha,\bar {\bf q}''_\alpha))
\tilde G_{{\sf F}\alpha}({\bf p}''_\alpha,\bar {\bf q}''_\alpha;z)
(-\bar V_\beta(\bar {\bf q}''_\beta,\bar {\bf q}'''_\beta))
\nonumber\\[2mm]
&\times&
G_{{\sf K}\beta}({\bf p}'_\beta,\bar {\bf q}'''_\beta;z)
(-\bar V_\gamma(\bar {\bf q}''_\gamma,{\bf q}_\gamma))
G_{{\sf K}\gamma}({\bf p}_\gamma,{\bf q}_\gamma;z)
\en }
\textit{\eq
U^{c,2}_{3\alpha\beta\gamma}({\bf p}''_\alpha{\bf q}''_\alpha;{\bf p}'_\beta;
{\bf p}_\gamma{\bf q}_\gamma)
&=&
G_{{\sf K}\alpha}({\bf p}''_\alpha,{\bf q}''_\alpha;z)
(-\bar V_\alpha({\bf q}''_\alpha,\bar {\bf q}''_\alpha))
\tilde G_{{\sf F}\alpha}({\bf p}''_\alpha,\bar {\bf q}''_\alpha;z)
(-\bar V_\beta(\bar {\bf q}''_\beta,\bar {\bf q}'''_\beta))
\nonumber\\[2mm]
&\times&
G_{{\sf K}\beta}({\bf p}'_\beta,\bar {\bf q}'''_\beta;z)
(-\bar V_\gamma(\bar {\bf q}''_\gamma,{\bf q}_\gamma))
G_{{\sf K}\gamma}({\bf p}_\gamma,{\bf q}_\gamma;z)
\nonumber\\[2mm]
&+&G_{{\sf K}\alpha}({\bf p}''_\alpha,{\bf q}''_\alpha;z)
(-\bar V_\alpha({\bf q}''_\alpha,\bar {\bf q}''_\alpha))
G_{{\sf K}\beta}({\bf p}'_\beta,\bar {\bf q}''_\beta;z)
(-\bar V_\beta(\bar {\bf q}''_\beta,\bar {\bf q}'''_\beta))
\nonumber\\[2mm]
&\times&
\tilde G_{{\sf F}\beta}({\bf p}'_\beta,\bar {\bf q}'''_\beta;z)
(-\bar V_\gamma(\bar {\bf q}''_\gamma,{\bf q}_\gamma))
G_{{\sf K}\gamma}({\bf p}_\gamma,{\bf q}_\gamma;z)
\nonumber\\[2mm]
&+&G_{{\sf K}\alpha}({\bf p}''_\alpha,{\bf q}''_\alpha;z)
(-\bar V_\alpha({\bf q}''_\alpha,\bar {\bf q}''_\alpha))
\tilde G_{{\sf F}\beta}({\bf p}'_\beta,\bar {\bf q}''_\beta;z)
(-\bar V_\beta(\bar {\bf q}''_\beta,\bar {\bf q}'''_\beta))
\nonumber\\[2mm]
&\times&
\tilde G_{{\sf F}\beta}({\bf p}'_\beta,\bar {\bf q}'''_\beta;z)
(-\bar V_\gamma(\bar {\bf q}''_\gamma,{\bf q}_\gamma))
G_{{\sf K}\gamma}({\bf p}_\gamma,{\bf q}_\gamma;z)
\en }
\textit{\eq
U^{c,3}_{3\alpha\beta\gamma}({\bf p}''_\alpha{\bf q}''_\alpha;{\bf p}'_\beta;
{\bf p}_\gamma{\bf q}_\gamma)
&=&
G_{{\sf K}\alpha}({\bf p}''_\alpha,{\bf q}''_\alpha;z)
(-\bar V_\alpha({\bf q}''_\alpha,\bar {\bf q}''_\alpha))
G_{{\sf K}\beta}({\bf p}'_\beta,\bar {\bf q}''_\beta;z)
(-\bar V_\beta(\bar {\bf q}''_\beta,\bar {\bf q}'''_\beta))
\nonumber\\[2mm]
&\times&
G_{{\sf K}\gamma}({\bf p}_\gamma,\bar{\bf q}''_\gamma;z)
(-\bar V_\gamma(\bar {\bf q}''_\gamma,{\bf q}_\gamma))
\tilde G_{{\sf F}\gamma}({\bf p}_\gamma,{\bf q}_\gamma;z)
\nonumber\\[2mm]
&+&
G_{{\sf K}\alpha}({\bf p}''_\alpha,{\bf q}''_\alpha;z)
(-\bar V_\alpha({\bf q}''_\alpha,\bar {\bf q}''_\alpha))
G_{{\sf K}\beta}({\bf p}'_\beta,\bar {\bf q}''_\beta;z)
(-\bar V_\beta(\bar {\bf q}''_\beta,\bar {\bf q}'''_\beta))
\nonumber\\[2mm]
&\times&
\tilde G_{{\sf F}\gamma}({\bf p}_\gamma,\bar{\bf q}''_\gamma;z)
(-\bar V_\gamma(\bar {\bf q}''_\gamma,{\bf q}_\gamma))
\tilde G_{{\sf F}\gamma}({\bf p}_\gamma,{\bf q}_\gamma;z)
\en }
\textit{\eq
U^{c,4}_{3\alpha\beta\gamma}({\bf p}''_\alpha{\bf q}''_\alpha;{\bf p}'_\beta;
{\bf p}_\gamma{\bf q}_\gamma)
&=&G_{{\sf K}\alpha}({\bf p}''_\alpha,{\bf q}''_\alpha;z)
(-\bar V_\alpha({\bf q}''_\alpha,\bar {\bf q}''_\alpha))
\tilde G_{{\sf F}\beta}({\bf p}'_\beta,\bar {\bf q}''_\beta;z)
\nonumber\\[2mm]
&\times&\dfrac{(-\bar V_\beta(\bar {\bf q}''_\beta,\bar {\bf q}'''_\beta))
(-\bar V_\gamma(\bar {\bf q}''_\gamma,{\bf q}_\gamma))}
{\biggl(M+\dfrac{({\bf p}_\gamma)^2}{2M_\gamma}
+\dfrac{(\bar {\bf q}''_\gamma)^2}{2\mu_\gamma}-z\biggr)}
\tilde G_{{\sf F}\gamma}({\bf p}_\gamma,{\bf q}_\gamma;z)
\en }
\textit{\eq
U^{c,5}_{3\alpha\beta\gamma}({\bf p}''_\alpha{\bf q}''_\alpha;{\bf p}'_\beta;
{\bf p}_\gamma{\bf q}_\gamma)
&=&\tilde G_{{\sf F}\alpha}({\bf p}''_\alpha,{\bf q}''_\alpha;z)
\dfrac{(-\bar V_\alpha({\bf q}''_\alpha,\bar {\bf q}''_\alpha))
(-\bar V_\beta(\bar {\bf q}''_\beta,\bar {\bf q}'''_\beta))}
{\biggl(M+\dfrac{({\bf p}'_\beta)^2}{2M_\beta}
+\dfrac{(\bar {\bf q}''_\beta)^2}{2\mu_\beta}-z\biggr)}
\nonumber\\[2mm]
&\times&
\tilde G_{{\sf F}\beta}({\bf p}'_\beta,\bar {\bf q}''_\beta;z)
(-\bar V_\gamma(\bar {\bf q}''_\gamma,{\bf q}_\gamma))
G_{{\sf K}\gamma}({\bf p}_\gamma,{\bf q}_\gamma;z)
\en }
\textit{\eq
&&U^{c,6}_{3\alpha\beta\gamma}({\bf p}''_\alpha{\bf q}''_\alpha;{\bf p}'_\beta;
{\bf p}_\gamma{\bf q}_\gamma)
=
\tilde G_{{\sf F}\alpha}({\bf p}''_\alpha,{\bf q}''_\alpha;z)
(-\bar V_\alpha({\bf q}''_\alpha,\bar {\bf q}''_\alpha))
G_{{\sf K}\alpha}({\bf p}''_\alpha,\bar {\bf q}''_\alpha;z)
(-\bar V_\beta(\bar {\bf q}''_\beta,\bar {\bf q}'''_\beta))
\nonumber\\[2mm]
&&\hspace*{2.cm}\times\,\,
G_{{\sf K}\gamma}({\bf p}_\gamma,\bar {\bf q}''_\gamma;z)
(-\bar V_\gamma(\bar {\bf q}''_\gamma,{\bf q}_\gamma))
\tilde G_{{\sf F}\gamma}({\bf p}_\gamma,{\bf q}_\gamma;z)
\nonumber\\[2mm]
&+&
\tilde G_{{\sf F}\alpha}({\bf p}''_\alpha,{\bf q}''_\alpha;z)
\dfrac{(-\bar V_\alpha({\bf q}''_\alpha,\bar {\bf q}''_\alpha))
(-\bar V_\beta(\bar {\bf q}''_\beta,\bar {\bf q}'''_\beta))}
{\biggl(M+\dfrac{({\bf p}'_\beta)^2}{2M_\beta}
+\dfrac{(\bar {\bf q}''_\beta)^2}{2\mu_\beta}-z\biggr)}
\tilde G_{{\sf F}\gamma}({\bf p}_\gamma,\bar {\bf q}''_\gamma;z)
(-\bar V_\gamma(\bar {\bf q}''_\gamma,{\bf q}_\gamma))
\tilde G_{{\sf F}\gamma}({\bf p}_\gamma,{\bf q}_\gamma;z)
\nonumber\\[2mm]
&+&\tilde G_{{\sf F}\alpha}({\bf p}''_\alpha,{\bf q}''_\alpha;z)
(-\bar V_\alpha({\bf q}''_\alpha,\bar {\bf q}''_\alpha))
\tilde G_{{\sf F}\alpha}({\bf p}''_\alpha,\bar {\bf q}''_\alpha;z)
\dfrac{(-\bar V_\beta(\bar {\bf q}''_\beta,\bar {\bf q}'''_\beta))
(-\bar V_\gamma(\bar {\bf q}''_\gamma,{\bf q}_\gamma))}
{\biggl(M+\dfrac{({\bf p}_\gamma)^2}{2M_\gamma}
+\dfrac{(\bar {\bf q}''_\gamma)^2}{2\mu_\gamma}-z\biggr)}
\tilde G_{{\sf F}\gamma}({\bf p}_\gamma,{\bf q}_\gamma;z)
\nonumber\\
&&
\en }
where
\textit{\eq
\bar {\bf q}''_\alpha&=&{\bf p}'_\beta+\dfrac{m_\beta}{M-m_\alpha}\,{\bf p}''_\alpha\, ,
\nonumber\\[2mm]
 \bar {\bf q}''_\beta&=&-{\bf p}_\alpha''-\dfrac{m_\alpha}{M-m_\beta}\,{\bf p}'_\beta\, ,
\nonumber\\[2mm]
 \bar {\bf q}'''_\beta&=&-{\bf p}_\gamma-\dfrac{m_\gamma}{M-m_\beta}\,{\bf p}'_\beta\, ,
\nonumber\\[2mm]
\bar {\bf q}''_\gamma&=&{\bf p}'_\beta+\dfrac{m_\beta}{M-m_\gamma}\,{\bf p}_\gamma \, .
\en }

It is straightforward to observe that the above terms reproduce the multiple-scattering series of the Faddeev equations in a finite volume (see section~\ref{sec:faddeev}) up to \textit{$O(\bar V_\alpha^3)$}.

\setcounter{figure}{0}
\setcounter{table}{0}
\setcounter{equation}{0}
\section{Product of two energy denominators}
\label{app:relation}

Consider the expression
\textit{\eq\label{eq:J}
J=\dfrac{1}{L^3}\sum_{\bf p}\dfrac{\phi({\bf p})}{(a^2-({\bf p}+{\bf c}_1)^2)
(b^2-({\bf p}+{\bf c}_2)^2)}
\doteq\dfrac{1}{L^3}\sum_{\bf p}\phi({\bf p})d_1d_2\, ,
\en }
where \textit{$\phi({\bf p})$} denotes a regular function. 

If \textit{${\bf c}_1={\bf c}_2={\bf 0}$} and \textit{$a^2\neq b^2$}, the proof of the desired relation
given in eq.~(\ref{eq:relation}) immediately follows from the identity
\textit{\eq
\dfrac{1}{(a^2-{\bf p}^2)(b^2-{\bf p}^2)}=
\dfrac{1}{b^2-a^2}\,
\biggl(\dfrac{1}{a^2-{\bf p}^2}-\dfrac{1}{b^2-{\bf p}^2}\biggr)
\en }
by using the splitting from eq.~(\ref{eq:formally}) in the individual terms
on the r.h.s. of this equation.

In the generic case, let us consider the partial-wave expansion
\textit{\eq
\phi({\bf p})d_2&=&\dfrac{\phi({\bf p}+{\bf c}_1-{\bf c}_1)}
{b^2-({\bf p}+{\bf c}_1-({\bf c}_1-{\bf c}_2))^2}\doteq
\dfrac{\phi_1({\bf p}+{\bf c}_1)}
{b^2-({\bf p}+{\bf c}_1-({\bf c}_1-{\bf c}_2))^2}
\nonumber\\[2mm]
&=&
\sum_{lm}{\cal Y}_{lm}({\bf p}+{\bf c}_1)
{\cal Y}^*_{lm}({\bf c}_1-{\bf c}_2)K_{1l}(({\bf p}+{\bf c}_1)^2)\, ,
\en }
and, analogously,
\textit{\eq
\phi({\bf p})d_1=\sum_{lm}{\cal Y}_{lm}({\bf p}+{\bf c}_2)
{\cal Y}^*_{lm}({\bf c}_1-{\bf c}_2)K_{2l}(({\bf p}+{\bf c}_2)^2)\, .
\en }
Further, define the quantities
\textit{\eq
\overline{\phi_1d_2}&=&\sum_{lm}{\cal Y}_{lm}({\bf p}+{\bf c}_1)
{\cal Y}^*_{lm}({\bf c}_1-{\bf c}_2)K_{1l}(a^2)\, ,
\nonumber\\[2mm]
\overline{\phi_2d_1}&=&\sum_{lm}{\cal Y}_{lm}({\bf p}+{\bf c}_2)
{\cal Y}^*_{lm}({\bf c}_1-{\bf c}_2)K_{2l}(b^2)
\en }
Then, the quantity \textit{$J$} from eq.~(\ref{eq:J}) can be rewritten as
\textit{\eq\label{eq:square}
J=\dfrac{1}{L^3}\sum_{\bf p}\biggl\{\biggl[
\phi d_1d_2-\theta_1d_1\overline{\phi_1d_2}-\theta_2d_2\overline{\phi_2d_1}\biggr]
+\theta_1d_1\overline{\phi_1d_2}+\theta_2d_2\overline{\phi_2d_1}\biggr\}\, ,
\nonumber\\[2mm]
\theta_1=f(a^2/\mu^2)\theta(\Lambda^2-({\bf p}+{\bf c}_1)^2)\, ,\quad\quad
\theta_2=f(b^2/\mu^2)\theta(\Lambda^2-({\bf p}+{\bf c}_2)^2)\, .
\en }
If
\textit{\eq
\hspace*{-.4cm}&&a^2>0\, ,\quad b^2>0\, ,\nonumber\\[2mm]
\hspace*{-.4cm}&&|(a^2-b^2)- ({\bf c}_1-{\bf c}_2)^2|>2|a|\,|{\bf c}_1-{\bf c}_2|,\quad\quad|(a^2-b^2)+ ({\bf c}_1-{\bf c}_2)^2|>2|b|\,|{\bf c}_1-{\bf c}_2|\, ,
\en }
then the quantities \textit{$\overline{\phi_1d_2},~\overline{\phi_2d_1}$} are 
non-singular and one may use the regular summation theorem. Namely, the
expression in the square brackets in eq.~(\ref{eq:square}) is non-singular,
so the summation can be replaced by integration there. 
Using the same technique as in section~\ref{sec:twobody}, one straightforwardly
arrives at eq.~(\ref{eq:relation}). 
Finally, the relation for the generic values of
the parameters \textit{$a,b,{\bf c}_1,{\bf c}_2$} can be obtained by 
analytic continuation
of both sides of eq.~(\ref{eq:relation}) in these parameters.


\begin{thebibliography}{99}

\bibitem{Isgur1}
  N.~Isgur and G.~Karl,
  Phys.\ Lett.\  B {\bf 72} (1977) 109.
  
  
\bibitem{Isgur2}
  N.~Isgur and G.~Karl,
  Phys.\ Rev.\  D {\bf 19} (1979) 2653
  [Erratum-ibid.\  D {\bf 23} (1981) 817].
  
\bibitem{hybrid1}
  Z.~P.~Li, V.~Burkert and Z.~J.~Li,
  Phys.\ Rev.\  D {\bf 46} (1992) 70.

  
  
  \bibitem{hybrid2}
   C.~E.~Carlson and N.~C.~Mukhopadhyay,
  Phys.\ Rev.\ Lett.\  {\bf 67} (1991) 3745.

\bibitem{breather}
  P.~A.~M.~Guichon,
  Phys.\ Lett.\  B {\bf 164} (1985) 361.

\bibitem{fivequark}
  O.~Krehl, C.~Hanhart, S.~Krewald and J.~Speth,
  Phys.\ Rev.\  C {\bf 62} (2000) 025207
  [arXiv:nucl-th/9911080].

\bibitem{ulf1}
  I.~Zahed, U.-G.~Mei{\ss}ner and U.~B.~Kaulfuss,
  Nucl.\ Phys.\  A {\bf 426} (1984) 525.



\bibitem{ulf2}
  U.-G.~Mei{\ss}ner and J.~W.~Durso,
  Nucl.\ Phys.\  A {\bf 430} (1984) 670.




\bibitem{Gockeler:2001db}
  M.~Gockeler, R.~Horsley, D.~Pleiter, P.~E.~L.~Rakow, G.~Schierholz, C.~M.~Maynard and D.~G.~Richards
                  [QCDSF Collaboration and UKQCD Collaboration and LHPC
                  Collaboration],
  Phys.\ Lett.\  B {\bf 532} (2002) 63
  [arXiv:hep-lat/0106022].



\bibitem{Melnitchouk:2002eg}
  W.~Melnitchouk {\it et al.},
  Phys.\ Rev.\  D {\bf 67} (2003) 114506
  [arXiv:hep-lat/0202022].


\bibitem{Lee:2002gn1}
  F.~X.~Lee and D.~B.~Leinweber,
  Nucl.\ Phys.\ Proc.\ Suppl.\  {\bf 73} (1999) 258
  [arXiv:hep-lat/9809095].

 
  
  \bibitem{Lee:2002gn2}
  
  F.~X.~Lee, S.~J.~Dong, T.~Draper, I.~Horvath, K.~F.~Liu, N.~Mathur and J.~B.~Zhang,
  Nucl.\ Phys.\ Proc.\ Suppl.\  {\bf 119} (2003) 296
  [arXiv:hep-lat/0208070].

\bibitem{Edwards:2003cd}
  R.~G.~Edwards, U.~M.~Heller and D.~G.~Richards  [LHP Collaboration],
  Nucl.\ Phys.\ Proc.\ Suppl.\  {\bf 119} (2003) 305
  [arXiv:hep-lat/0303004].


\bibitem{F.Lee}
  N.~Mathur {\it et al.},
  Phys.\ Lett.\  B {\bf 605} (2005) 137
  [arXiv:hep-ph/0306199].



\bibitem{Sasaki:2003xc1}
  S.~Sasaki, T.~Blum and S.~Ohta,
  Phys.\ Rev.\ D {\bf 65} (2002) 074503
  [hep-lat/0102010].
  
  \bibitem{Sasaki:2003xc2}
   S.~Sasaki,
  Prog.\ Theor.\ Phys.\ Suppl.\  {\bf 151} (2003) 143
  [arXiv:nucl-th/0305014].

 
  
  \bibitem{Sasaki:2003xc3}
  K.~Sasaki, S.~Sasaki and T.~Hatsuda,
  Phys.\ Lett.\ B {\bf 623} (2005) 208
  [hep-lat/0504020].


\bibitem{Basak:2007kj}
  S.~Basak {\it et al.},
  Phys.\ Rev.\  D {\bf 76} (2007) 074504
  [arXiv:0709.0008 [hep-lat]].

\bibitem{Cohen:2009zk}
  S.~Cohen {\it et al.},
  PoS {\bf LAT2009} (2009) 112
  [arXiv:0911.3373 [hep-lat]].

\bibitem{Bulava:2010yg1}
  J.~Bulava {\it et al.},
  Phys.\ Rev.\  D {\bf 79} (2009) 034505
  [arXiv:0901.0027 [hep-lat]].
  
  
  \bibitem{Bulava:2010yg2}
  J.~Bulava {\it et al.},
  Phys.\ Rev.\  D {\bf 82} (2010) 014507
  [arXiv:1004.5072 [hep-lat]].

\bibitem{Engel:2010my}
  G.~P.~Engel, C.~B.~Lang, M.~Limmer, D.~M\"ohler and A.~Sch\"afer  [BGR [Bern-
                  Graz-Regensburg] Collaboration],
  Phys.\ Rev.\  D {\bf 82} (2010) 034505
  [arXiv:1005.1748 [hep-lat]].


\bibitem{Mahbub:2010rm}
  M.~S.~Mahbub, W.~Kamleh, D.~B.~Leinweber, P.~J.~Moran and A.~G.~Williams
                  [CSSM Lattice collaboration],
  arXiv:1011.5724 [hep-lat].

\bibitem{Lin:2011da}
  H.~W.~Lin and S.~D.~Cohen,
  arXiv:1108.2528 [hep-lat].


\bibitem{Lin:2011ti}
  H.~W.~Lin,
  Chin.\ J.\ Phys.\  {\bf 49} (2011) 827
  [arXiv:1106.1608 [hep-lat]].



\bibitem{Borasoy:2006fk}
  B.~Borasoy, P.~C.~Bruns, U.-G.~Mei{\ss}ner and R.~Lewis,
  Phys.\ Lett.\  B {\bf 641} (2006) 294
  [arXiv:hep-lat/0608001].


\bibitem{luescher-torus}
  M.~L\"uscher,
  Nucl.\ Phys.\  B {\bf 354} (1991) 531.


\bibitem{rho1}
  M.~Gockeler, R.~Horsley, Y.~Nakamura, D.~Pleiter, P.~E.~L.~Rakow, G.~Schierholz and J.~Zanotti
                  [QCDSF Collaboration],
  PoS {\bf LATTICE2008} (2008) 136
  [arXiv:0810.5337 [hep-lat]].


 \bibitem{rho2}
 S.~Aoki {\it et al.}  [CS Collaboration],
  Phys.\ Rev.\  D {\bf 84} (2011) 094505
  [arXiv:1106.5365 [hep-lat]].

 
\bibitem{lage-KN}
  M.~Lage, U.-G.~Mei{\ss}ner and A.~Rusetsky,
  Phys.\ Lett.\  B {\bf 681} (2009) 439
  [arXiv:0905.0069 [hep-lat]].
\bibitem{He}
  C.~Liu, X.~Feng and S.~He,
  Int.\ J.\ Mod.\ Phys.\  A {\bf 21} (2006) 847
  [arXiv:hep-lat/0508022].



\bibitem{scalar}
  V.~Bernard, M.~Lage, U.-G.~Mei{\ss}ner and A.~Rusetsky,
  JHEP {\bf 1101} (2011) 019\\{}
  [arXiv:1010.6018 [hep-lat]].


\bibitem{oset}
  M.~D\"oring, U.-G.~Mei{\ss}ner, E.~Oset and A.~Rusetsky,
 Eur.\ Phys.\ J.\  A {\bf 47} (2011) 139
  [arXiv:1107.3988 [hep-lat]].



\bibitem{Torres:2011pr}
  A.~M.~Torres, L.~R.~Dai, C.~Koren, D.~Jido and E.~Oset,
  arXiv:1109.0396 [hep-lat].

\bibitem{kappa}
  M.~D\"oring and U.-G.~Mei{\ss}ner,
  arXiv:1111.0616 [hep-lat].
\bibitem{Belyaev}
  V.~B.~Belyaev,
  ``Lectures On The Theory Of Few Body Systems,''
{\it  Berlin, Germany: Springer (1990) 134 p. (Springer series in nuclear and particle physics)}


\bibitem{Luu}
  T.~Luu and M.~J.~Savage,
  Phys.\ Rev.\  D {\bf 83} (2011) 114508
  [arXiv:1101.3347 [hep-lat]].


\bibitem{Lage-distributions}
  V.~Bernard, M.~Lage, U.-G.~Mei{\ss}ner and A.~Rusetsky,
  JHEP {\bf 0808} (2008) 024\\{}
  [arXiv:0806.4495 [hep-lat]].

\bibitem{Doring:2011ip}
  M.~D\"oring, J.~Haidenbauer, U.-G.~Mei{\ss}ner and A.~Rusetsky,
  Eur.\ Phys.\ J.\  A {\bf 47} (2011) 163
  [arXiv:1108.0676 [hep-lat]].



\bibitem{luescher-2.}
  M.~L\"uscher,
  Commun.\ Math.\ Phys.\  {\bf 105} (1986) 153 (1986).

\bibitem{Kowalski-3}
  K.~L.~Kowalski,
  Phys.\ Rev.\  D {\bf 7} (1973) 1806;
  Nucl.\ Phys.\  A {\bf 264} (1976) 173.

\bibitem{Manning}
I. Manning,   Phys.\ Rev.\  D {\bf 5} (1972) 1472.

\bibitem{Kowalski-K}
  K.~L.~Kowalski,
  Phys.\ Rev.\  D {\bf 5} (1972) 395.

\bibitem{Doring:2009yv1}
  M.~D\"oring, C.~Hanhart, F.~Huang, S.~Krewald and U.-G.~Mei{\ss}ner,
  Nucl.\ Phys.\  A {\bf 829} (2009) 170
  [arXiv:0903.4337 [nucl-th]].
 
  
  \bibitem{Doring:2009yv2}
  M.~D\"oring, C.~Hanhart, F.~Huang, S.~Krewald and U.-G.~Mei{\ss}ner,
  Phys.\ Lett.\  B {\bf 681} (2009) 26
  [arXiv:0903.1781 [nucl-th]].


\bibitem{Kreuzer:2008bi}
  S.~Kreuzer and H.~W.~Hammer,
  Phys.\ Lett.\  B {\bf 673} (2009) 260
  [arXiv:0811.0159 [nucl-th]].



\bibitem{Kreuzer:2009jp}
  S.~Kreuzer and H.~W.~Hammer,
  Eur.\ Phys.\ J.\  A {\bf 43} (2010) 229
  [arXiv:0910.2191 [nucl-th]].



\bibitem{Kreuzer:2010ti}
  S.~Kreuzer and H.~W.~Hammer,
  Phys.\ Lett.\  B {\bf 694} (2011) 424
  [arXiv:1008.4499 [hep-lat]].


\bibitem{Bour:2011ef}
  S.~Bour, S.~K\"onig, D.~Lee, H.~W.~Hammer and U.-G.~Mei{\ss}ner,
  Phys.\ Rev.\  D {\bf 84} (2011) 091503
  [arXiv:1107.1272 [nucl-th]].

\bibitem{Gradsteyn}
I.S. Gradshteyn and I.M. Ryzhik, 
Table of Integrals, Series and Products, Academic Press (2007).

\end{thebibliography}
\end{document}